\documentclass[a4paper,11pt]{article}
\usepackage{graphicx} 
\usepackage{orcidlink,thumbpdf,lmodern}
\usepackage[authoryear, comma, numbers, sort]{natbib}
\usepackage{lineno}

\usepackage{subcaption}
\usepackage{threeparttable}
\usepackage{float}
\usepackage{multirow}
\usepackage{multicol}
\usepackage{color}
\usepackage{xspace}
\usepackage{enumerate}
\usepackage{listings}
\usepackage{makecell}

\usepackage[paper=a4paper, top=1.5cm, bottom=1.5cm, left=2cm, right=2cm] {geometry}

\newcommand{\code}[1]{\textit{#1}'}

\newcommand{\fct}[1]{\code{#1()}}
\newcommand{\pkg}[1]{\code{#1()}}
\newcommand{\proglang}[1]{\code{#1()}}

\begin{document}



\title{Flexible Approach for Statistical Disclosure Control\\ in Geospatial Data}

\author{ Jon Olav Skøien$^a$~\orcidlink{0000-0002-8706-1986}, Nicolas Lampach$^{b,c}$~\orcidlink{0009-0002-5099-7147}, Helena Ramos$^b$, Rudolf Seljak$^d$,\\  Renate Koeble$^a$, Linda See$^e$~\orcidlink{0000-0002-2665-7065} and Marijn van der Velde$^f$~\orcidlink{0000-0002-9103-7081}\\
$^{a}$\small\emph{ARHS Developments, Luxembourg}\\[-5pt]
$^{b}$\small\emph{European Commission, Eurostat E.1., Luxembourg}\\[-5pt]
$^{c}$\small\emph{Institute of Sustainable Economic Development, BOKU University, Austria}\\[-5pt]
$^{d}$\small\emph{Private consultant in the field of official statistics, Slovenia}\\[-5pt]
$^{d}$\small\emph{International Institute for Applied Systems Analysis (IIASA), Austria}\\[-5pt]
$^{e}$\small\emph{European Commission, Joint Research Centre D.5., Italy}}

\date{}
\maketitle
\thispagestyle{empty}


\begin{abstract}
\normalsize 
We develop a flexible approach by combining the Quadtree-based method with suppression to maximize the utility of the grid data and simultaneously to reduce the risk of disclosing private information from individual units. To protect data confidentiality, we produce a high resolution grid from geo-reference data with a minimum size of 1 km nested in grids with increasingly larger resolution on the basis of statistical disclosure control methods (i.e threshold and concentration rule). While our implementation overcomes certain weaknesses of Quadtree-based method by accounting for irregularly distributed and relatively isolated marginal units, it also allows creating joint aggregation of several variables. The method is illustrated by relying on synthetic data of the Danish agricultural census 2020 for a set of key agricultural indicators, such as the number of agricultural holdings, the utilized agricultural area and the number of organic farms. We demonstrate the need to assess the reliability of indicators when using a sub-sample of synthetic data followed by an example that presents the same approach for generating a ratio (i.e., the share of organic farming). The methodology is provided as the open-source \textit{R}-package \textit{MRG} that is adaptable to use with other geo-referenced survey data underlying confidentiality or other privacy restrictions.\\[-10pt]

\end{abstract}



\newpage
\section{Introduction}
Across many domains it is common to distribute data in the form of grids, where the grid cells represent sums or averages of the recorded values within the cells. These grids usually have a common resolution for all grid cells, independent of the number of records. This works well for many applications, but there are cases where we cannot or do not want to disseminate grid cell values unless they respect certain restrictions. These can be based on confidentiality (we cannot reveal information that might lead to identification of individual records), statistical reliability (we do not want to reveal information with too high an uncertainty) or other, more field-dependent restrictions. 

Census and sample survey data are examples of data sets where any distribution of the data must respect both confidentiality and statistical reliability restrictions. Historically, the solution has been to adopt a very conservative and protective approach to the dissemination of data, where the aggregation level of the publicly available data is very coarse compared to the high detail level of the raw data. Although the methods and software presented here are applicable to any type of census and survey data, the focus of this paper is on the dissemination of data from the European agricultural census. 

An agricultural census involves the regular and systematic collection of data on the structure of a nation’s agricultural sector. The unit of data collection is the agricultural holding, which is comprised of the parcels of land and livestock managed by a single entity, such as an individual, household, or a public or private sector organization, for the purpose of agricultural production. By collecting information at regular intervals over time, such as the size of the holding, crop and livestock production and agricultural inputs, any changes in the agricultural sector can be monitored as well as their impacts on food security and the environment \citep{fao_world_2017v1}.

Decennial agricultural censuses have been taking place since 1930 as part of the World Agricultural Census \citep{ribi_forclaz_agriculture_2016}, which is an initiative that has been continued by the Food and Agriculture Organization of the United Nations (FAO) since 1950. \cite{fao_world_2017v1, fao_world_2017v2} provides countries with a recommended methodology that countries can adapt within their own monitoring systems, which includes the identification of a set of essential variables that should be collected to ensure global comparability. The guidance also includes different modes of operation from a traditional census every ten years to a more integrated program of censuses and surveys, where a sample survey is used to collect data during years in between the decennial census, as well as a modular approach, which is used to collect more detailed information on specific areas of interest. 

In the European Union (EU), a decennial agricultural census is undertaken across Member States (MS) along with a sample survey every 3 to 4 years, referred to as the Farm Structure Survey\footnote{The name originated from the former Regulation 1166/2008 on the European farm survey \citep{the_european_commission_regulation_2008} and most users are familiar with this term. Current legislation \citep{the_european_commission_regulation_2018} amended the name to Integrated Farm Statistics (IFS), which is used less frequently, and therefore we opt to refer to it as the FSS in this paper.} (FSS). Stipulated by Regulation (EU) 2018/1091 of the European Parliament and of the Council of 18 July 2018 on Integrated Farm Statistics \citep{the_european_commission_regulation_2018}, the collection of information in the FSS follows a common methodology in order to produce comparable and representative statistics across Member States and over time. In addition, EFTA countries Iceland, Switzerland, and Norway also participated in the 2020 census, and the FSS covered more than 9 million farm holdings. 

FSS data are used to assess the state of agriculture across the EU, monitoring trends and transitions in the structure of farms \citep{eurostat_farm_2024}. For example, \cite{neuenfeldt_explaining_2019} used FSS data to determine the drivers of farm structure change, finding that past farm structure explains the largest amount of variation but other drivers such as environmental conditions, prices, subsidies and income also play a role. The data are also key inputs to the management and evaluation of the Common Agricultural Policy (CAP) in terms of its environmental, economic and social impacts, and as inputs to CAP reforms, modelled, for example, using CAPRI \citep{barreiro-hurle_modelling_2021}. In addition to the CAP, FSS data are valuable for other policy areas, including the environment, climate change, employment and regional development (e.g., \cite{copus_study_2006}, \cite{einarsson_subnational_2020}).

FSS data are a form of microdata, which refers to any data that are collected from a respondent in a census or survey \citep{fao_world_2017v2}. Agricultural census and survey data are additionally complicated because agricultural holdings can contain information related to commercial operations or sensitive personal data. The release of census and survey data are, therefore, subject to confidentiality legislation, which states that data about individuals or enterprises cannot be released or disclosed. Statistical disclosure control is the process by which national statistical offices ensure that any confidentiality legislation is applied \citep{fao_world_2017v2,eurostat2019}. Different methods of statistical disclosure are used including table redesign such that information in the tables is aggregated, cell suppression in which values are completely omitted, and adjustment of values using different approaches such as rounding, controlled adjustment to replace cells with `safe' values, and perturbation, where random noise is added to cells \citep{hundepool2010handbook,liu_statistical_2009, european_commission_statistical_2021,templ2017statistical,quatember2013family}.  

In the case of the FSS and to ensure that individual farms cannot be identified, the tables are first aggregated to coarse administrative levels (i.e., NUTS2, NUTS1 or even national level depending on the MS) before release by the EU's Statistical Office (Eurostat), curating FSS data for all Member Statses in the EU. However, there would be considerable value for policy design, policy impact assessment and scientific research more generally, in having access to data at a finer spatial resolution. Moreover, with advances in technology and the increasing trend to provide open access to government data across many sectors, new methods for disseminating data from censuses and surveys are needed \citep{shlomo_statistical_2018}.

In this paper we present a methodology (implemented with a bespoke \textit{R} package) that  takes data collected at individual level, implements legal confidentiality rules, and produces aggregate values for a multi-resolution spatial grid. The method can also apply a contextual suppression, where some grid cells with few records are suppressed if it means that neighboring grid cells can be disseminated with a high resolution. We demonstrate the approach using the variables Utilized Agricultural Area (UAA) and Organic Utilized Agricultural Area (UAA\_ORG) for the  country of Denmark. Although the methodology is specifically demonstrated using FSS data, such an approach could also be adapted for releasing other individual census and survey based data that are subject to legal rules of disclosure, or where a certain reliability is demanded for each grid cell. The method is released in the \textit{R}-package \textit{MRG} on the Comprehensive R Archive Network (CRAN) to make the methodology available to other applications. The functionality has been developed with flexibility so that different restrictions than those relevant to the FSS data can be easily added.  

\section{Data} 

To provide a more detailed overview of the European survey on the structure of agricultural holdings, we first present the data collection framework and then describe the detailed topics and variables in the database. We also provide a synthetic data set for the Danish 2020 agricultural census along with the \textit{R}-package, hands-on examples and guidance to produce the maps. In addition, we outline the confidentiality rules and quality assessment of the indicators that are implemented in the methodology that produces a high resolution grid of the data.

\subsection{European surveys on the structure of agricultural holdings}
European surveys on the structure of agricultural holdings have been carried out since 1966, and they aim to provide statistical knowledge for the monitoring and evaluation of related policies, in particular the CAP as well as environmental, climate change adaptation and land use policies. To reduce the burden on national administrations, Regulation (EU) 2018/1091 on integrated farm statistics provides a new framework by distinguishing between core and module variables\footnote{ The complete list and description of variables surveyed during the European agricultural census 2020 can be found in the Implementing Regulation (EU) 2018/1874 of 29 November 2018 on the data to be provided for 2020 under Regulation (EU) 2018/1091 of the European Parliament and of the Council on integrated farm statistics and repealing Regulations (EC) No 1166/2008 and (EU) No 1337/2011.}, which vary in frequency and representativeness \citep{the_european_commission_regulation_2018}. It is required that the information on the core variables (e.g, general structural agricultural variables) should cover 98\% of the utilised agricultural area and 98\% of the livestock units of each MS. The modules contain information on specific topics such as the labour force, animal housing or irrigation, and can be carried out on samples of agricultural holdings by meeting the precision requirement laid down in Annex V of Regulation (EU) 2018/1091. 

\subsubsection{The raw survey data}
National data providers (i.e., national statistical institutes, ministries of agriculture or other governmental bodies) prepare the questionnaire, conduct the interviews and complete the survey with additional information from administrative registers (e.g., wine, bovines, integrated information and the control system). The individual records at farm level are encrypted and transmitted to Eurostat via a secure system that implements an automated procedure to validate the content and structure of the micro data. For the first time during the 2020 agricultural census, Eurostat introduced an automated error detection procedure, leading to higher quality statistics. 

\begin{table}[t!]
    \centering
\resizebox{0.9\textwidth}{!}{
    \begin{tabular}{llrrrr}
    \hline \hline\\[-5pt]
    Year & Type & Variables & Surveyed farms* (MM) & Population covered* (MM) & Countries \\      
    \hline\\[-5pt]
    2010   & Census  & 419 & 12.81 & 13.03 & 33\\
    2013   & Sample  & 358 & 1.73  & 11.04 & 30\\
    2016   & Sample  & 363 & 1.69  & 10.55 & 30\\
    2020   & Census  & 364 & 9.03  & 9.16  & 30\\
    \hline \hline\\[-25pt]
    \multicolumn{6}{l}{\multirow{2}{=}{\footnotesize \textit{Note.} *Covers all Member States, candidate and EFTA countries for the respective data collection year.}}\\\\[-10pt]
    \multicolumn{6}{l}{\footnotesize Further details about the coverage can be found in \cite{Eurostat2023_coverage}.}
    \end{tabular}}
    \caption{Data collection overview of the farm structure survey}
    \label{tab:data_coll_fss}
\end{table}

While an agricultural census is carried out every 10 years, sample surveys are administered during interim years. Table \ref{tab:data_coll_fss} summarises the data collection for the last decade by highlighting the number of variables, the number of surveyed farms, the population covered and the number of countries participating in the survey rounds. A substantial volume of information was collected during the 2020 survey campaign, which was comprised of more than 300 variables from around 9.03 million agricultural holdings. In sample survey years such as 2016, 1.69 million agricultural holdings were surveyed, which at that time represented approximately 10.55 million holdings. It is worth mentioning that the lower sample numbers will give lower accuracy and quality of estimates from sample data compared to the agricultural census. Therefore, we have also introduced a reliability criterion for the indicators used in the production of the multi-resolution grid data which will also ensure comparability. 

\subsubsection{Synthetic data}
For practical purposes, we have derived a synthetic data set from the original 2020 agricultural census micro data to avoid any malicious disclosure of sensitive information. Although synthetic data sets are a feasible way to provide public access to the data by mitigating any confidentiality concerns, there have only been a few attempts made to create synthetic public files of micro data collected by official statistical institutes \citep{burgard2017synthetic,drechsler2012new,patyk2020modelling,speth2022synthetic,taub2020impact,templ2017simulation,wimmer2023note}. 

\cite{rubin1993statistical} and \cite{little1993statistical} have both proposed a framework to create synthetic data by mirroring the information from the original records. Statistical methods to compute synthetic data range from general regression, Bayesian inference, non-parametric methods of regression and classification tree approaches, with the aim of preserving the underlying joint distribution of the original data. 

We applied a hot-deck procedure - originally developed to impute missing information - to substitute a data entry from the original data (i.e., the recipient) by using a value from a similar record (i.e., the donor) within the same classification group \citep{andridge2010review,ford1983overview,joenssen_hot_2012}. Unlike other methods, the synthetic data generated by this hot-deck approach contains only plausible values. 

A single hot deck imputed data set is computed for each country individually. First, records are partitioned into homogeneous groups so that the donors follow the same distribution as the recipients. Data points from the recipients are substituted sequentially based on a value from a varying pool of donors. Furthermore, the nearest neighbour matching technique using distance metrics is applied to select the most appropriate donor from the pool of donors. For a few of the discrete variables, such as $FARMTYPE$, $SO\_EUR$, $HLD\_FEF$ and $NUTS2$, a donor was chosen randomly by preserving the original empirical distribution or they were simply randomly decoded (i.e., renamed). The variable containing information about the geographical location ($GEO\_LCT$) of the agricultural holding was imputed by restricting the donor to the same country. To assess the quality rating system (i.e., the reliability), we created an artificial sample ($SAMPLE$) with the respective extrapolation factors ($EXT\_MODULE$) based on stratification. The sample size consists of approximately one third of the synthetic 2020 census for Denmark. 

The empirical distribution of the two main variables of interest of the synthetic data, $UAA$ and $UAAXK0000\_ORG$ are widely preserved within the different economic size classes (see Figure \ref{fig:syn_uaa} and \ref{fig:syn_uaax} in Appendix \ref{app:synth}). Table \ref{tab:syn} in \ref{app:synth} also confirms that there is no statistical differences for both variables between the synthetic and the original data file. 

The synthetic data set is attached to the developed \proglang{R} package. Table \ref{tab:synthetic} lists and explains the variables that are included in the synthetic data set. The value of the synthetic data lies in illustrating the methodology, the implicit trade-offs between the spatial resolution and disclosure, and the variable specific considerations, and it should also ease the uptake and implementation of the \proglang{R} package by interested institutions. 
\begin{table}[t!]
     \centering
    \resizebox{\textwidth}{!}{
    \begin{tabular}{l l l}
      \hline \hline\\[-5pt]
     Variable code & Variable description & Type  \\
       \hline\\[-5pt]
    $COUNTRY$  & ISO code of the country name & discrete\\
    $YEAR$     & Survey/census year & continous\\
    $REGIONS$    & NUTS-2 region & discrete\\
    $GEO\_LCT$  & Geographical location of the farm & discrete\\
    $HLD\_FEF$  & Main frame$^{*}$ (HLD\_FEF=0) or frame extension (HLD\_FEF=1) & dummy \\
     \multirow[t]{2}{*}{$STRA\_ID\_CORE$} & Stratification ID (any positive integer number). Stratification is mainly applied based & \multirow[t]{2}{*}{continuous}\\
     & on farm type, standard output and land size & \\
    $SAMPLE$    & Artificial sample created based on stratification (1=Holding is included and 0=excluded) & binary\\
    \multirow[t]{2}{*}{$EXT\_CORE$} & Extrapolation factor of Core (principally the value is 1, but it  & \multirow[t]{2}{*}{continuous}\\
    & can vary according to non-response adjustment or calibration) &\\
    \multirow[t]{2}{*}{$EXT\_MODULE$} & Extrapolation factor related to the artificial sample & continuous\\
    \multirow[t]{3}{*}{$FARMTYPE$}  & Typology of the farm. This code list contains the types of farms described by their activities  & \multirow[t]{3}{*}{discrete}\\
    & (e.g., raising cattle, cultivating arable crops). Farms are classified into different types  & \\
    & according to their dominant activity: crops, livestock and mixed-farming (e.g FT1\_SO) & \\
    $SO\_EUR$  & Classes of standard output (e.g KE\_0, KE\_LT2, KE2-5,\dots, KE\_GE500) & discrete\\
    $UAA$  & Utilised agricultural area of the holding & continuous\\
    $UAAXK0000\_ORG$  & Organic utilised agricultural area, excluding kitchen gardens & continuous\\
      \hline \hline\\[-5pt]
    \end{tabular}}
    \begin{tablenotes}
      \scriptsize
      \item [*] *Art.3 of Regulation 2018/1091 lays down the minimum requirement of data coverage (98\% of the total UAA (excluding kitchen gardens) and 98\% of the livestock units of each Member State.). Where the frame does not represent the minimum requirements, Member States shall extend the frame (frame extension) by establishing lower thresholds \citep{the_european_commission_regulation_2018}.
    \end{tablenotes}
    \caption{Overview of the variables selected for the synthetic data set}
    \label{tab:synthetic}
\end{table}

\subsection{Disclosure control and quality rating} \label{section:DisclosureControl}

Official statistics are governed by a fundamental principle that protects the confidentiality of individuals or organisations and produces high-quality official statistics by masking sensitive information according to international and European law\footnote{Separate national laws (EU/EEA/EFTA) might contain stricter (or laxer) rules related to the disclosure of personal information.} \citep{the_european_commission_regulation_2009,the_european_commission_regulation_2018,eurostat2019,trewin2007principles}. At a higher spatial resolution, there is a legally binding obligation to employ appropriate aggregation and disclosure control to render spatial data sets accessible to the public \citep{the_european_commission_regulation_2018}. Furthermore, the Implementing Regulation (EU) 2018/1874 defines a set of rules for disclosing information from European surveys on the structure of agricultural holdings collected at farm location including the use of the 1 km INSPIRE Statistical Units Grid for pan-European data. In addition to the standard rules for tabular data, a key requirement is that values can only be disseminated at a 1 km grid when the cell includes more than \textbf{ten} agricultural holdings. Alternatively, aggregating to a nested 5 km or larger grid size is required to satisfy the aforementioned requirement \citep{the_european_commission_commission_2020}.

A disclosure occurs when an intruder correctly finds or determines some values about an individual or organization from the data released. \cite{duncan1989risk} differentiate between two types of disclosure risk: identity disclosure and attribute disclosure. While the former occurs when a record can be directly linked to an individual, the latter refers to the knowledge gained about an individual or organization from the attribute(s) in the data released. Statistical Disclosure Control (SDC) techniques are widely deployed to reduce the risk of disclosing private information at an acceptable level, while maximising the utility of the data \citep{quatember2013family,templ2017statistical}. From the two broad families of methods that exist, the perturbative method modifies the data prior to publication by adding random noise such as rounding to the nearest multiple of ten. Non-perturbative techniques reduce the amount of information by suppressing or aggregating the data. The optimal mixture of SDC should strike a balance between the mandatory privacy protection of the statistical output and the accessibility to the data at the highest available spatial resolution \citep{quatember2013family}. 

In agreement with member states, \cite{Eurostat2020IFSmanual} has provided a series of recommendations in the confidentiality charter for disclosure control. For the dissemination of aggregated tabular statistics, key elements include mandatory compliance with the threshold\footnote{Suppression of cells representing less than four agricultural holdings.} and dominance rule\footnote{Suppression of cells when one or two contributors are dominant.}, and the statistical output must satisfy certain quality criteria\footnote{Data accuracy is evaluated based on sampling errors that can be estimated from the sample itself using the standard errors of the estimated values. If the coefficient of variation of the estimated values is larger than 35\%, the cell is usually suppressed.}. These rules also apply to the dissemination of spatially explicit data on a grid. 

Another important aspect that is receiving increasing attention is second order confidentiality, which occurs when the value of a suppressed sensitive cell can be determined from neighbouring cells or from other publicly available sources. In terms of gridded data, it is possible that cells become identifiable when both high-resolution gridded data and low-resolution NUTS data are published. Applying gap-filling methods to both data sets to impute the suppressed values would put the disclosure of private information at risk \citep{higgins_atmospheric_2012}. This threat can be overcome by carefully choosing the size of the grid cells and the type of administrative regions for the dissemination of the data.

Lastly, a quality control of the gridded value is necessary. This is usually not an issue when (almost) the entire population has been sampled, as in census years, but in sample years, the use of larger extrapolation weights will introduce prediction errors to the gridded values. The prediction errors can be estimated as a function of the sample size, population size, sampled values and possible stratification. The Integrated Farm Statistics Manual \citep[Section 4.6]{Eurostat2023IFSmanual} also requires that the relative standard error (coefficient of variation) of the estimate should be less than 0.35, as otherwise, the information cannot be disclosed.

\section{Methods}
\subsection{Multi-grid approach} 
Several different methods can, with different advantages and disadvantages, be used to create a gridded data set that respects the confidentiality rules above. The following subsections will present some of the possibilities, including our proposed method for FSS data.

\subsubsection{Gridding}
From a point data set like the FSS data, an unlimited number of regular grids can be created. For the methods below, we first need to create a set of base grids of different resolutions. The first three resolutions are specifically mentioned in the EU regulations, which require these to be 1, 5 and 10 km \citep{the_european_commission_commission_2020}. There are no restrictions on coarser resolution grids. However, the methodologies below require the grids to have a hierarchical structure where the coarser resolution grids must be integer multiples of the higher resolution grids. For example, coarser resolution grids could be 10, 20, 40, 80 and 160 km, or 10, 50 and 100 km. However, 10, 20, 50 and 100 km would not be possible as the 50 km grid includes 2.5 grid cells (in each direction) from the 20 km grid and is, therefore, not an integer multiple.

If the data are to be disseminated as a regular grid, the confidentiality rules must be examined for each different grid level, and the final resolution will be the highest resolution at which the confidentiality rules are respected for all grid cells. This method is intuitive and is quick to implement. It will work well when the data are fairly well distributed over the domain of interest. However, if the density of holdings is considerably lower in some regions where only coarse resolution grids respect the confidentiality rules, then the resolution will need to be equally coarse in the rest of the data set. 

\begin{figure}[t!]
    \centering
    \includegraphics[width=0.85\linewidth]{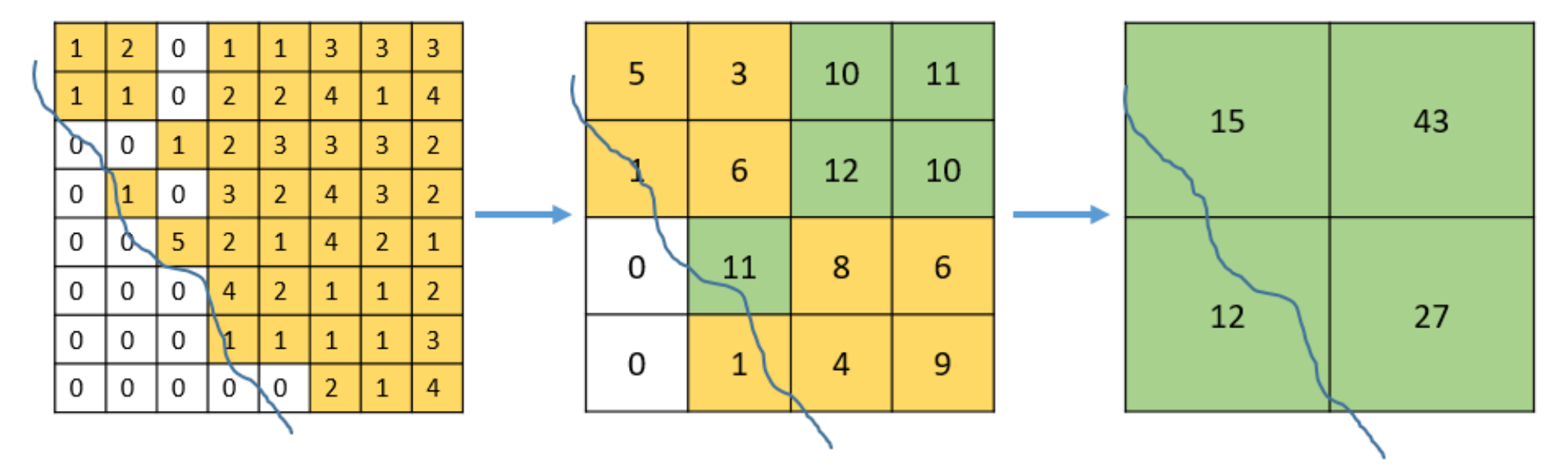}
    \includegraphics[width=0.13\linewidth]{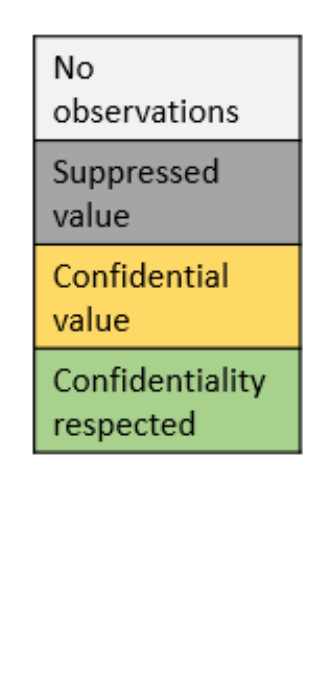}
    \caption{Example of gridded data, moving from a higher to a coarser resolution. The numbers represent the number of holdings per grid cell. The line represents a border or coastline. }
    \label{fig:gridding}
\end{figure}

Figure \ref{fig:gridding} shows a fictitious example in which the number of holdings have first been aggregated in 2*2 blocks to a lower resolution grid. Assuming that 10 holdings are necessary for disclosing the information from a grid cell, none of the grid cells will pass the confidentiality rules. In the second grid, 4*4 blocks of the original cells have been aggregated to larger cells. Here several of the cells (green) respect the confidentiality rules, but others (yellow) do not, so this grid cannot be disclosed either. In the last grid in Figure \ref{fig:gridding}, all grid cells respect the confidentiality rules. However, we can see that the data in the upper right grid cell has been aggregated to a much coarser resolution than necessary, and hence this solution is not optimal on its own.

\subsubsection{Value suppression}
\begin{figure}[t!]
    \centering
    \includegraphics[width=0.85\linewidth]{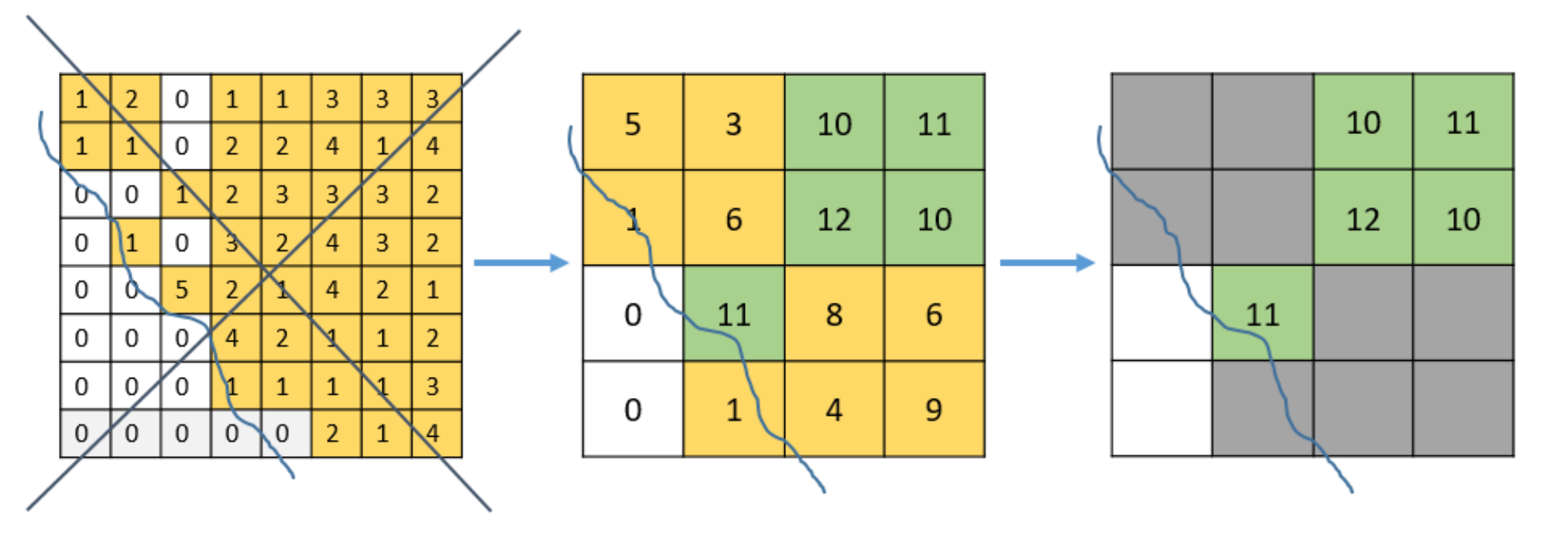}
    \includegraphics[width=0.13\linewidth]{figures/legend1.pdf}
    \caption{Example of suppression of grid cell values that do not respect the confidentiality rules. The highest resolution grid cannot be used, as all values would have been suppressed with a limit of 10.}
    \label{fig:suppression}
\end{figure}

Another relatively simple approach for producing a grid that respects the confidentiality rules is to suppress the values from grid cells where the confidentiality rules are not respected. Figure \ref{fig:suppression} shows an example using the same fictitious data. The leftmost panel of Figure \ref{fig:suppression} shows a situation where all grid cells would be suppressed so no data could be released. The middle panel shows an aggregation of the original grid to a suitable resolution as a first step. The confidentiality rules are applied to determine which cells should be suppressed and the final result is seen in the right panel of Figure \ref{fig:suppression}. As a result, a relatively large number of cells become empty. This can happen if there is a large difference in the density of holdings in different regions of the study domain. Another issue with this method is that the gridded sum of the holdings will be considerably less than the actual number of holdings, as several values have been removed. Hence, this solution is also not adequate for disseminating high resolution FSS data.

\subsubsection{Reallocation of grid cell values}

\begin{figure}[t!]
    \centering
    \includegraphics[width=0.85\linewidth]{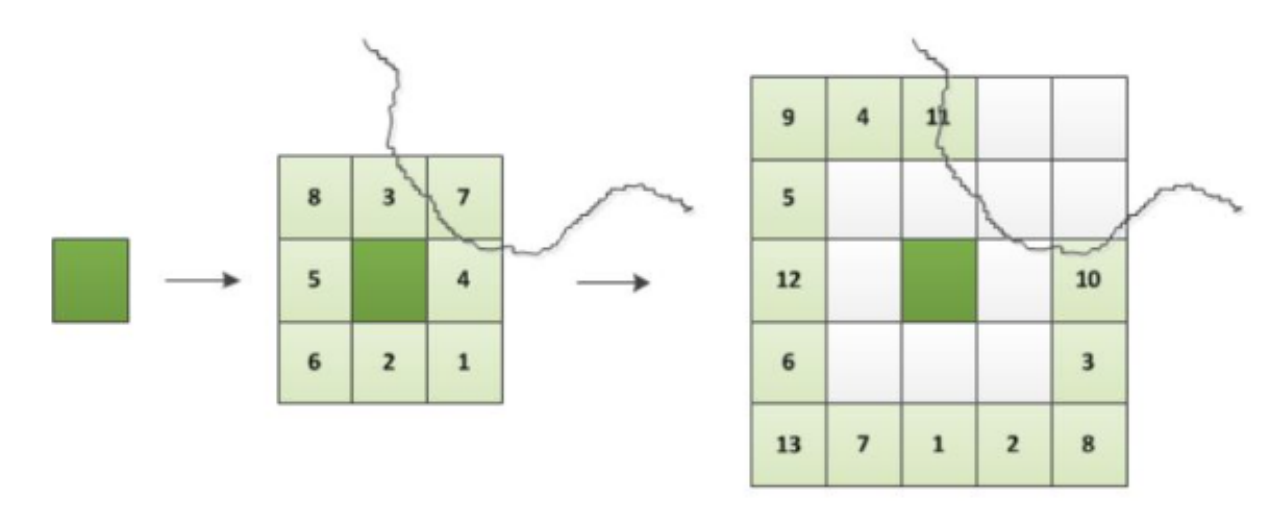}
    \caption{Example showing how the value in one cell can be attributed to one of the neighbouring cells. The dark green shading indicates the cell that does not pass the confidentiality rules, light green indicates cells with values that do, while white cells do not contain any holdings. The line indicates a NUTS3 boundary. In the middle graph, a value can be randomly reallocated to another grid cell within the same NUTS3 boundary. In the graph on the far right, no neighbouring cells directly adjacent contain at least 1 holding so the search is extended to neighbouring values in a larger search radius.  Source: \cite{Eurostat2020IFSmanual}.}
    \label{fig:swapping0}
\end{figure}

A third method would be to reallocate grid cell values that do not respect the confidentiality rules to one of the neighbouring grid cells. This can be implemented in different ways. The Integrated Farm Statistics Manual \citep{Eurostat2020IFSmanual} recommends that the values from a grid cell that do not pass the confidentiality rules should be reallocated (added) to a random neighbouring cell within the same NUTS3 region containing at least one holding. If there are no such neighbouring grid cells, the search radius is extended, as shown in Figure \ref{fig:swapping0}. 

\begin{figure}[t!]
    \centering
    \includegraphics[width=0.85\linewidth]{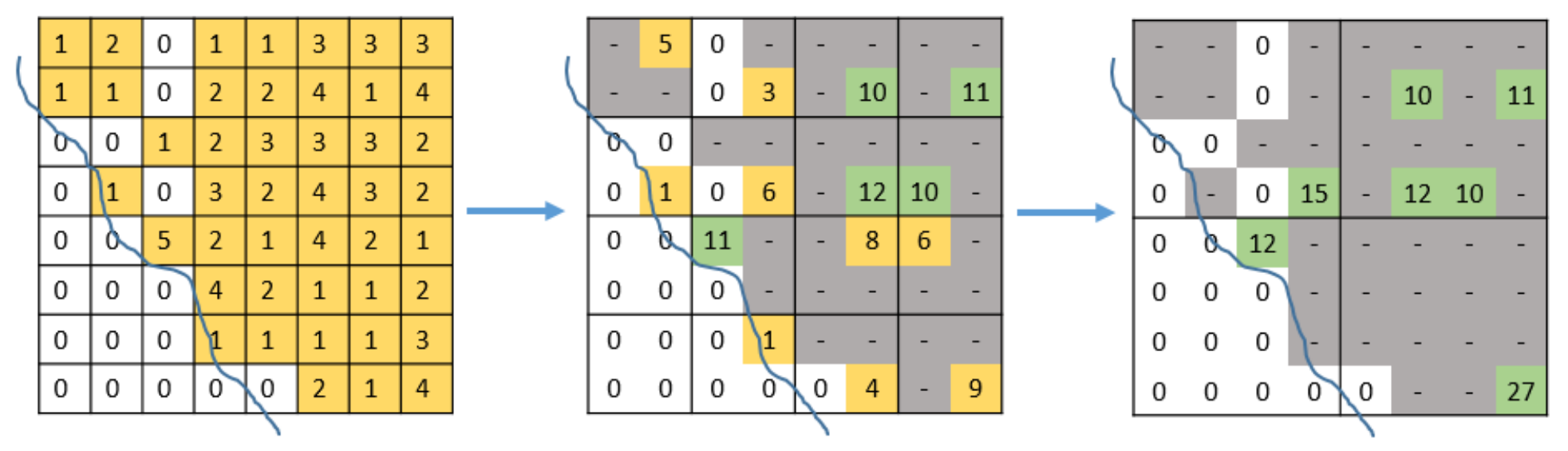}
    \includegraphics[width=0.12\linewidth]{figures/legend1.pdf}
    \caption{Example showing how values are hierarchically reallocated from cells that do not respect the confidentiality rules to neighbouring cells in the same 2*2 block. The line shows a NUTS3 boundary.}
    \label{fig:swapping1}
\end{figure}

The \proglang{R} package described in this paper includes a modification to this recommended method, where it is possible to reallocate cell values within the same 2*2 blocks that are used in the hierarchical grids described above. The value of the cells that do not respect the confidentiality rules are reallocated to one of the grid cells in the block that already respect the confidentiality rules, or to the grid cell with the highest value if none of them respects the confidentiality rule. This approach is demonstrated in Figure \ref{fig:swapping1}. In the leftmost figure, all values are below the threshold in the region of interest (shown with yellow shading). The lowest values are reallocated to the grid cell with the highest value in each 2*2 block, resulting in the central panel. Here the remaining green grid cells in the upper right corner respect the confidentiality rules in addition to one grid cell in the lower left quadrant. In the rightmost panel, the remaining values that did not respect the confidentiality rules have been reallocated to a neighbouring grid cell in a 2*2 block of the intermediate grid, resulting in more than 10 holdings in each displayed cell.

\begin{figure}[t!]
    \centering
    \includegraphics[width=0.85\linewidth]{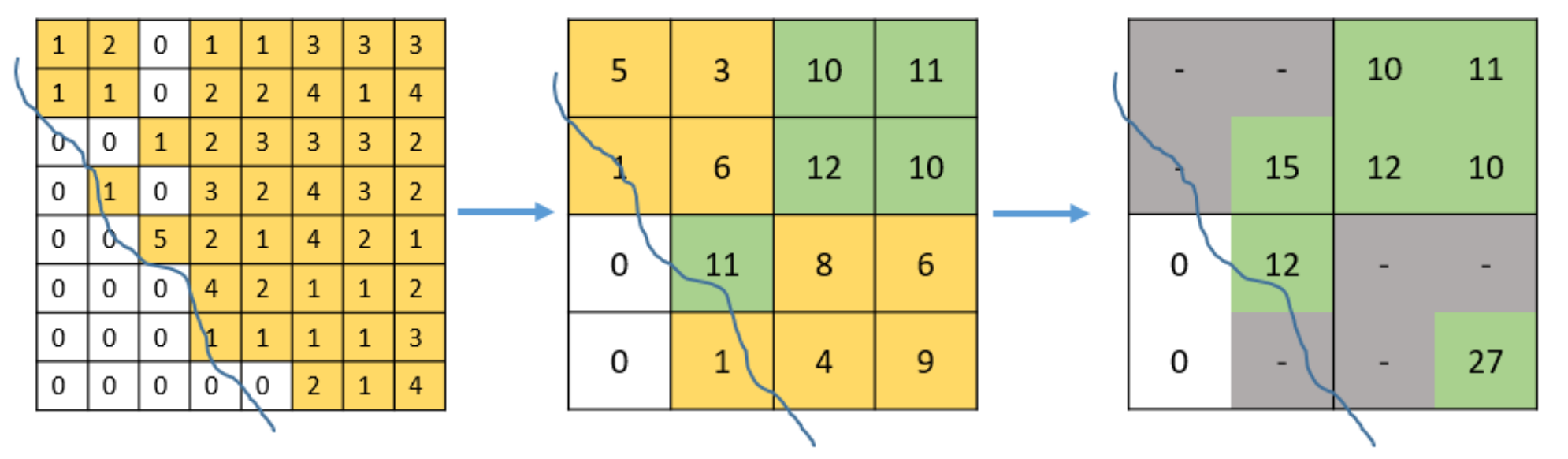}
    \includegraphics[width=0.12\linewidth]{figures/legend1.pdf}
    \caption{Example showing how values are first aggregated to a coarser resolution before the values from cells that do not respect the confidentiality rules are reallocated to cells with higher values within the same block. The line shows a NUTS3 boundary.}
    \label{fig:swapping2}
\end{figure}

A disadvantage with this method is that it can result in many empty grid cells if the resolution chosen at the start is too high. Depending on the pattern of the densities of holdings in the region of interest, it could be better to start with a slightly coarser resolution grid. This is demonstrated in Figure \ref{fig:swapping2}, where the grid is aggregated to a coarser resolution before reallocating the values. The values are identical to the ones shown in Figure \ref{fig:swapping1}, but they refer to larger (and more realistic) cell sizes. 

\subsubsection{Multi-resolution grid}

\begin{figure}[t!]
    \centering
    \includegraphics[width=0.85\linewidth]{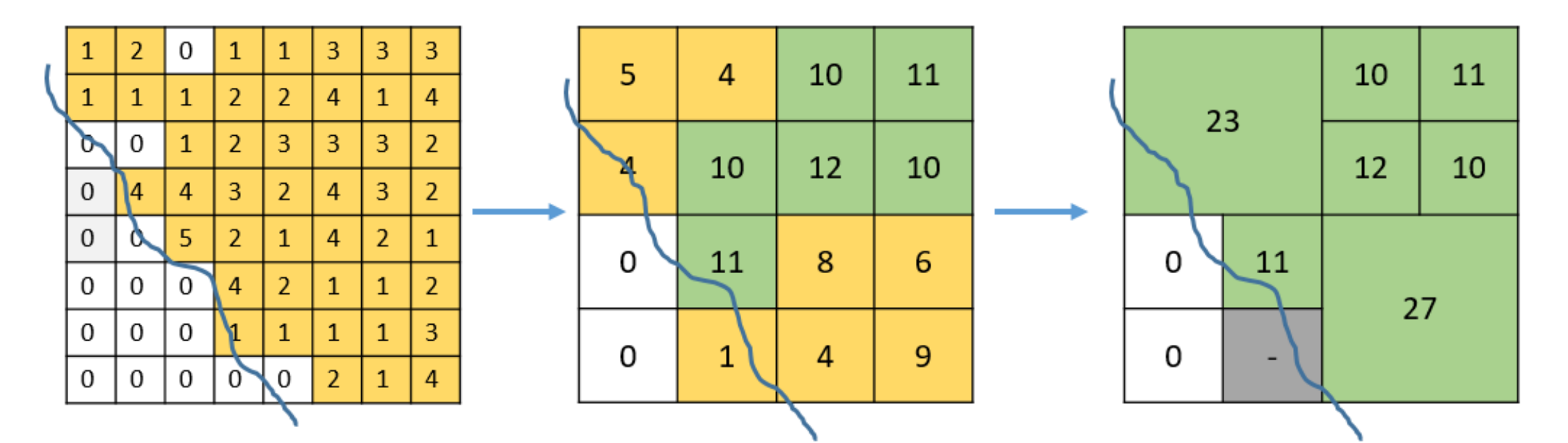}
    \includegraphics[width=0.12\linewidth]{figures/legend1.pdf}
    \caption{Example of a multi-resolution grid, moving from a higher to a coarser resolution. The numbers represent the number of holdings per grid cell. The line shows a border effect (regional border or land/sea).}
    \label{fig:multires_1}
\end{figure}

Another approach is to disclose information with a variable grid size that has a hierarchical data structure, also referred to as multi-resolution grid or quadtree \citep{Asim2023EQmultigrid, Behnisch2013quadtree, Eurostat2020IFSmanual,lagonigro_aquadtree_2020}. The idea here is that the resolution of the grid will vary according to the local density of the observations to ensure that the confidentiality rules are respected for all grid cells. 
An example of this is shown in Figure \ref{fig:multires_1} with the same fictitious data and the same aggregation (middle panel) as in the value suppression approach. However, when reducing the resolution towards the third and final map in Figure  \ref{fig:multires_1}, the four cells in the upper right corner are not aggregated, as they already respect the confidentiality rules. Hence it is possible to share the data with a higher resolution in this area and at a coarser resolution in the rest of the map.

The method is sensitive to islands and borders, where it might be difficult to include a sufficient number of holdings, even for relatively large grid cells, when most of the grid cell does not include the data. Therefore, it might not be recommended to continue the process until absolutely all grid cells respect the confidentiality rules, but either to stop at a maximum resolution to suppress the values from grid cells that still do not respect the confidentiality rules, or to suppress some of the smaller grid cells that do not contribute much to the total value. Similar to the value suppression method, this means that the sum of the values from the grid cells will be lower than the total number of holdings. However, this need for suppression will affect considerably fewer grid cells than when using suppression on a regular grid. This method is, therefore, the one that has been chosen for the high resolution gridded data from the FSS.

Second order confidentiality, as outlined in section \ref{section:DisclosureControl}, could be an issue here, although to a lesser degree than the publication of grids at different resolutions, as the number of suppressed cells would be lower. Moreover, the rounding applied to the grids is an effective way to perturb the direct comparison of data published at different resolutions. 

There are some cases where aggregation might not be desired. In the situation where a relatively large single grid cell does not respect the confidentiality rules, it is fine to aggregate it if the neighbouring grid cells are also relatively large. However, it would be unfortunate if the single cell was aggregated with many smaller grid cells that could otherwise be disseminated at a high resolution. The added value of being able to present a value for a region with very few farms is perhaps lower than what is lost by having to aggregate to a lower resolution. We have, therefore, also introduced the possibility for the user to set a minimum share of a grid cell value relative to the possible lower resolution grid cell before it is necessary to aggregate. The check will either be against the variable of interest (UAA, livestock units, etc.) or the number of holdings if no variable is analysed. If the limit, for example, is 0.1, the grid cell with 1 in the lower left quadrant would not lead to aggregation in Figure \ref{fig:multires_1}, as it represents less than 10\% of the value of the possible lower resolution grid cell. Instead, it would be left as is and suppressed in the post-processing step.

\begin{figure}[t!]
    \centering
    \includegraphics[width=1.1\textwidth]{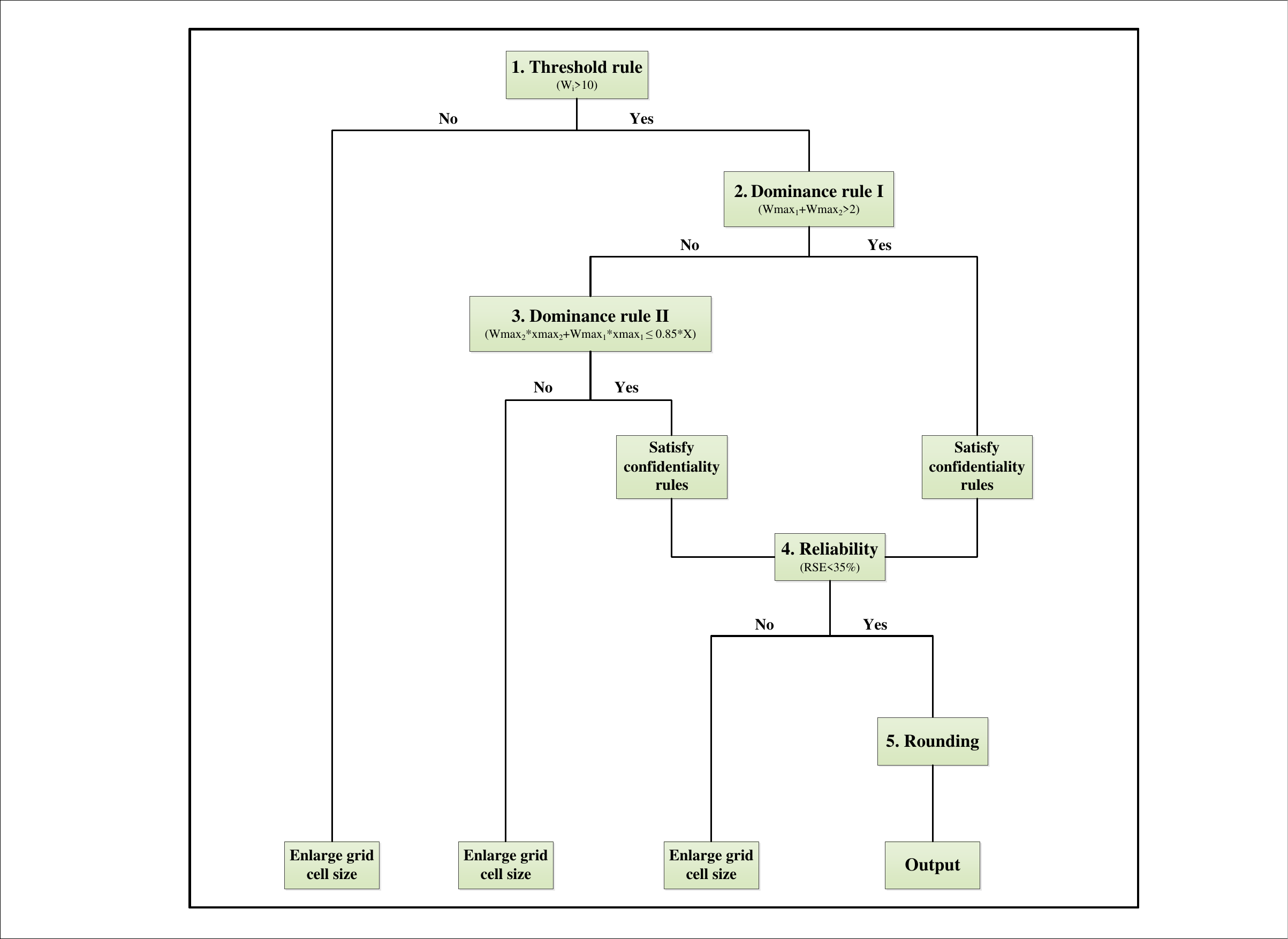}
    \caption{Flowchart showing the rules that are applied for the release of FSS data. Where the rules are not satisfied, the grid cell sizes must be increased as implemented in the multi-grid solution, (unless their impact is smaller than a certain limit).}
    \label{fig:stat_req}
\end{figure}

Figure \ref{fig:stat_req} displays the iterative process of producing a nested structure of multi-hierarchical grids satisfying a set of confidentiality rules and quality requirements. We denote the level of resolution $k \in K$ with $K=\{k_0,k_1,\dots,k_m\}$ where $k_0$ is the highest resolution (1 km for FSS) and $k_m$ the lowest resolution. The iteration starts with $i_1$, the possible aggregation from $k_0$ to $k_1$ and continues until reaching the maximum level $k_m$ ($i\in \{i_1,\dots,i_m\}$).

The confidentiality rules are evaluated in the following order, where the threshold rule must be passed, whereas it is sufficient to pass one of the dominance rules: 

\begin{enumerate}
    \item Threshold rule:
    
    If the aggregated extrapolated number of agricultural holdings in grid cell $l$ ($w_l$) for resolution $k_0$ in iteration $i_1$ is less than ten ($W_{l}<10$ with $W_l=\sum_{j=1}^{n_l} w_j$) where $n_l$ is the number of records in $l$, then the grid cell size must be enlarged to $k_1$ and the confidentiality rules for the new grid cell will be assessed in iteration $i_2$;
      \item Dominance rule I:
     
     If, after ordering the variable of interest in descending order, the sum of the weights $w_{j max_1}$ of the highest value $x_{max_{1}}$ ($W_{max_{1}}=w_{j max_1}\times x_{max_{1}}$) and of the second highest value $x_{max_{2}}$ ($W_{max_{2}}=w_{j max_2}\times x_{max_{2}}$) is greater than two ($W_{max_{1}}+W_{max_{2}}>2$), then the dominance rules are satisfied for the grid cell at $k_0$ and the reliability of the results needs to be assessed. (Note that the weights are rounded before this step, so larger than 2 means at least 3). Otherwise Dominance rule II needs to be satisfied;
      \item Dominance rule II:
      
      If the two potential dominant contributors are less than or equal to 85\% of the extrapolated aggregated value of the grid cell ($W_{max_{2}}\times x_{max_{2}}+W_{max_{1}}\times x_{max_{1}}\leq0.85\times X$), then the confidentiality rules are satisfied; otherwise the grid cell size must be enlarged to $k_1$ and the confidentiality rules for the new grid cell will be assessed in iteration $i_2$;
  \item Reliability of the results:
  
  If the coefficient of variation (Relative Standard Error (RSE)) for the grid cell at $k_0$ is less than 35\%, then the indicator is reliable (to be disseminated with a warning if above 25\%); otherwise the grid cell size must be enlarged to $k_1$ and and the confidentiality rules for the new grid cell will be assessed in iteration $i_2$;
  \item Rounding:
  
  After the last iteration, and as a measure to add further perturbation to the disclosed information, all non-confidential extrapolated number of holdings and extrapolated aggregated values of variables are rounded to the nearest multiple of ten.
\end{enumerate}

\subsubsection{Joint aggregation} \label{section:jointAggregation}

Some indicators can be a function of two or more variables. A typical example would be the ratio between a variable and the utilized agricultural area (UAA) of a grid cell. If the numerator is Var1 and the denominator is Var2, both variables need to respect the confidentiality rules. If multi-grids are created for the two variables separately, the grids will most likely not match. Instead, it is possible to grid the variables jointly, where high resolution grid cells are replaced with low resolution grid cells for one of the variables if necessary. After creating a common grid, it is then possible to do the division. There is also a possibility for creating a second grid with the same resolution as a previous grid. This will work fine if the new variable has a similar or higher density as the first one, but will give a high number of suppressed grid cells if the density is lower.

\subsubsection{Post-processing of the data}

Two different methods of post-processing can be applied after creating the multi-resolution grids as described above. First, the procedures above do not ensure that the confidentiality rules are respected when the lowest resolution has been reached, and when some grid cells are left unaggregated because they are adjacent to many high resolution grid cells. Thus, there could still be some grid cells that do not respect the confidentiality rules.  The values in cells that do not pass the confidentiality rules can be suppressed or the values can be reallocated to neighbouring grid cells. However, it is not clear how this neighbourhood reallocation should be done. For this reason, the first approach is chosen. 

\subsection{Implementation of the approach} 
The method has been implemented as a package in the statistical environment \proglang{R} \citep{r_core_team_r_2024}. The package contains a range of functions that help the user to create grids that respect the confidentiality rules required when releasing FSS data or other types of survey/census data. Similar functionality (i.e., the multi-resolution grid) is implemented in the packages \pkg{AQuadtree} \citep{lagonigro_aquadtree_2020} and \pkg{sdcSpatial \citep{jonge_sdcSpatial_2022}} . However, these packages lack the flexibility to include the dominance rule, and they have no support for several of the other features in this package.

The functionality in the \pkg{MRG} package uses methods from the \pkg{sf} package \citep{pebesma_simple_2018,pebesma_spatial_2023}, spatial analysis functionality from the packages \pkg{stars} \citep{pebesma_spatial_2023},  \pkg{terra} \citep{hijmans_terra_2024} and \pkg{vardpoor} \citep{breidaks_vardpoor_2020}, and some features from \pkg{plyr} \citep{wickham_plyr_2011}, \pkg{dplyr} \citep{wickham_dplyr_2023}, \pkg{magrittr} \citep{bache_magrittr_2022}, \pkg{purrr} \citep{wickham_purrr_2023}, \pkg{sjmisc} \citep{ludecke_sjmisc_2018}, and \pkg{rlang} \citep{henry_rlang_2023}.

The package is released on the Comprehensive R Archive Network (CRAN: https://cran.r-project.org/) under the GPL(>=3) License.

\begin{figure}[t!]
    \centering
    \includegraphics[width=\textwidth,trim=10 10 10 10,clip]{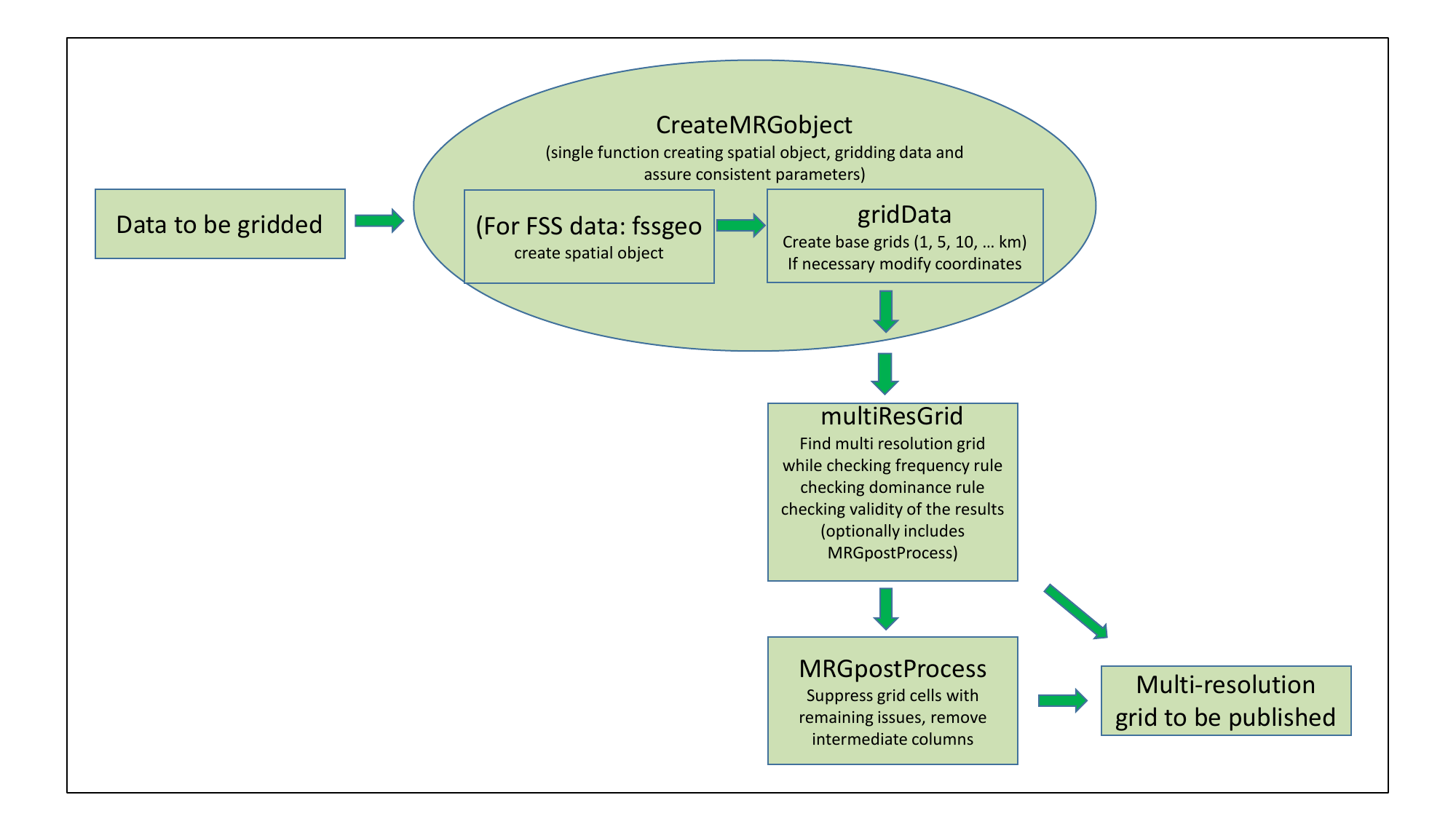}
    \caption{Flowchart of the procedure to produce the multi-resolution grid}
    \label{fig:flowchart2}
\end{figure}

The gridding procedure, which is shown in Figure \ref{fig:flowchart2}, is implemented as follows: 
\begin{enumerate}
  \item Read the data set (as a csv, Excel or other file format) containing the variable to be gridded.
  \item The data set has to be converted to a spatial \code{sf}-type object \citep{pebesma_simple_2018}. The spatial information in the FSS data has a special format, encoded as a string containing the country name, the coordinate reference system, the resolution and the coordinates. The \fct{fssgeo} function will parse this string and create the spatial object to be used in further analyses. For other types of census and survey data, users will need to create an \code{sf}-object themselves based on how the data are spatially represented.
  \item Create a gridded data set of one or more variable(s) with different resolutions, either using the \fct{gridList} function or the \fct{createMRGobject} function. The difference between these functions is that the first one only creates a set of grids, whereas the second one will create an object including both the grids and all other necessary information, in an object, making it easier to ensure that the same variables and parameters are submitted to different functions. The \fct{createMRGobject}-function will also call \fct{fssgeo} if the data seems to be of FSS-type.
  
  The default resolutions (of 1, 5, 10, 20, 40, 80 km) follow the regulations up to 10 km, and aggregates of 2*2 grid cells for lower resolutions, but the user can change these. If only the number of holdings is of interest, then no variable name is required for this function. 
  For all other variables (e.g., the UAA, the number of livestock, etc.), the column name(s) should be a parameter of this function. If there is a weight associated with the variable (if some observations in the data are samples from a larger group), the column name(s) with the weights should also be added. If only one weight column is given, this will be applied to all variables. Otherwise, one weight should be given for each variable. 
  
  Problems will arise if the observations fall exactly on the border between grid cells since it will not be clear which cell they belong to. This is the case for FSS data, where the coordinates have been mapped to the corners of a 1 km grid. If using \fct{gridList}, the coordinates should be adjusted before calling the function (for example, with \fct{st\_jitter} or \fct{locAdjFun}). The same modified coordinates will then also have to be submitted to the \fct{multiResGrid} function. If creating an $MRG$-object, the adjustment of the coordinates can instead be done through a variable $locAdj$. The function will then  adjust the locations to shift them away from the borders. The value can either be one of $LL$ (lower left), $LR$ (lower right), $UL$ (upper left), or $UR$ (upper right), where the value refers to where in a grid cell the coordinate is located. The default for FSS data is that the coordinate should be from the lower left corner ($LL$).  Alternatively, the value can be $jitter$, which means that the function will add a random jitter to the coordinates,  randomly distributing the holdings to either side of a border.   
  \item Create the multi-resolution grid using the function \fct{multiResGrid}. Note that value reallocation can be done with the function \fct{remSmall} but this approach is not used in the examples provided here. Most of the parameters for this function were mentioned in the methods section, with default values reflecting the standards for the FSS data. The parameters should be included as a part of the $FSS$-object if \fct{createMRGobject} was used above. Some of the parameters not already mentioned include:
  \begin{itemize}
      \item the choice of whether to apply the confidence rules to all variables individually (\code{confrules = "individual"}), or to just look at the first variable,
      \item the parameter $suppresslim$, which indicates if grid cells with a value less than the $suppresslim$ share of an aggregated grid cell should be left unaggregated, rather to be suppressed at a later stage, 
      \item the possibility to add another function (\fct{userfun}) that tests other criteria that might be necessary for alternative applications of the method (see more details below), and
      \item a logical variable $postProcess$, which indicates if post processing should already take place in this function (the default is $TRUE$), or if the user wants to examine the raw data before post processing in a separate function.    
  \end{itemize}
  
  \item If not already done as part of the \fct{multiResGrid} function, run the \fct{MRGpostProcessing} function, which will check that all grid cells respect the confidentiality rules and suppress values from those cells that do not. Additionally, this function will round the variables according to the rounding rule, where the default is rounding to the closest 10.

\end{enumerate}

The procedure above applies to one or more variables from a data set that are gridded together. A special case is the need for a gridded ratio between variables (as discussed in the section \emph{Joint Aggregation}. For example, this could be a variable that is expressed in relation to the number of farms or the total UAA in a grid cell. If both variables have already been gridded together, the ratio between them can be created by dividing the variables. If for some reason one of the grids has already been created and cannot be recomputed, it is possible to create a similar grid of the second variable by passing the first grid as the first parameter (\code{gdl}) to  the function \fct{multiResGrid}. This will work well if the density of the second variable is similar or higher than the first variable, but will create a grid with a high number of suppressed values of the density is lower.

Although \fct{gridList} was presented above as a way of creating the hierarchical grids, it is probably easier to create an $MRG$-object with the function \fct{createMRGobject}. This should include all the necessary data such as the original data set, the resolutions, the parameters for confidentiality rules, etc. This can then be used as an input to \fct{multiResGrid}, instead of having to specify all the different parameters every time. \fct{MRGpostProcess} does not use the \code{MRG}-object, but the necessary parameters will be included as attributes to the output of \fct{multResGrid}. This will together ensure that the same data set and values are used consistently throughout the entire procedure. 


There is also an additional feature in the package, where a user-defined function  \fct{userFun} can be passed to the gridding function in addition to the existing confidentiality rules. This could be useful if a user wants to use the package for creating a multi-resolution grid, but with different rules than the ones implemented.
A simple example for the FSS data could be to restrict some of the variables in a grid cell to a maximum value. FSS data are reported at the administrative location, which in some cases can be in a municipality centre rather than the location of the parcels. In those cases, the \fct{userFUn} could flag a grid cell for aggregation if the total UAA of the grid cell is more than the size of the grid cell itself. 
The input of the function must be similar to the functions for the confidentiality rules. Further details about this are given in the help files for the multi-resolution grid function. 


\section{Results}

A synthetic data set and a subset of it were used to demonstrate the procedure at a country level. The first is the synthetic data for Denmark, representing an agricultural census like the FSS while the second is a subset of the previous data set, representing an agricultural survey that would be collected in between census years and therefore has less data. The procedure was then applied to actual data on UAA from the 2020 FSS for all of Europe followed by a demonstration using LUCAS data \citep{Ballin2022} to produce a multi-resolution grid of landscape features.  

\subsection{Gridded FSS data} 

\begin{figure}[t!]
    \centering
    \includegraphics[width=\textwidth]{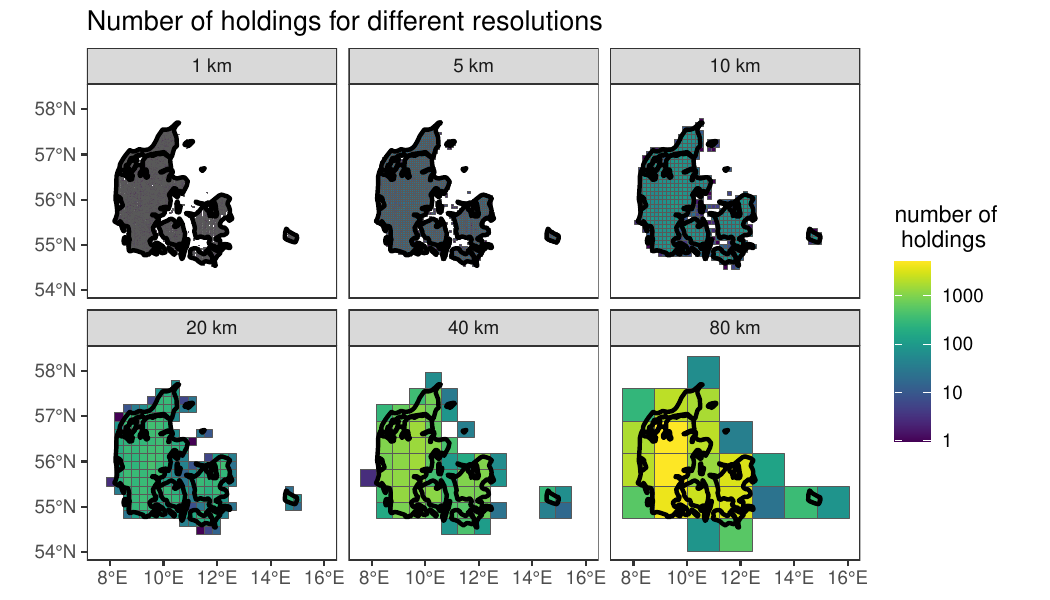}
    \caption{Number of holdings per grid cell for different grid cell sizes for Denmark (synthetic data) }
    \label{fig:gridDenmark}
\end{figure}

The gridded data set with different resolutions can be seen as the base for the more advanced methods below. Figure \ref{fig:gridDenmark} shows the number of holdings per grid cell for different grid cell sizes for Denmark: 1, 5, 10, 20, 40 and 80 km, using the synthetic data set described above. In this first example, we will only look at the number of holdings per grid cell, and use the frequency rule (i.e., minimum of 10 holdings in a grid cell) as the only confidentiality rule. We cannot see much in the 1 km grid, but none of the grid cells have more than the confidentiality limit of 10 holdings. It is still difficult to visualize the individual grid cells in the 5 km grid, but we can see that a large share of them are now above the limit of 10 holdings. Around 80\% of the approximately 2000 grid cells have more than 10 holdings. For the 10, 20, 40 and 80 km grids, there are 52, 11, 1 and 0 grid cells with less than 10 holdings, respectively, meaning that the 80 km grid is the finest grid cell size at which it is not necessary to suppress values from any of the grid cells.  

The example data set is included in the package as $ifs_{dk}$. The function \fct{fssgeo} creates a spatial object and modifies the coordinates so they are not exactly on the grid lines. Then the \fct{gridList} function creates the a list of base grids that is used to create the multi-resolution grid later. $ress$ is the set of grid cell sizes that we would like to use in the gridding procedure. The complete scripts for reproducing the plots and numeric information are given in the Supplementary Material.  

We can see that it is the coastal grid cells in Denmark, in particular, that make it necessary to continue aggregating until none of the grid cells are below the confidentiality limit. In the 40 and 80 km grids, we can see how some of the coastal grid cells are partly or mainly in the sea, resulting in the need to aggregate to such low resolutions for the entire data set. This example was presented in this manner for illustration purposes. However, in the rest of this section we will clip the grid cells to the coastlines, which produces a better visual representation of the agricultural activity.

\subsection{Multi-gridded FSS data} 

\subsubsection{Gridded number of farms with confidentiality rules} \label{section:GriddedFSSData}

With the set of grids from the previous step ($ifl$), we can run the \fct{multiResGrid} function to create a multi-resolution grid, which is based only on the number of holdings as a confidentiality rule. 

If we add the observations to the gridding procedure, it is also possible to apply the dominance rule, in this case based on the UAA. The dominance rule is used by default if the observations are provided. 

\begin{figure}
    \centering
    \includegraphics[width=\textwidth]{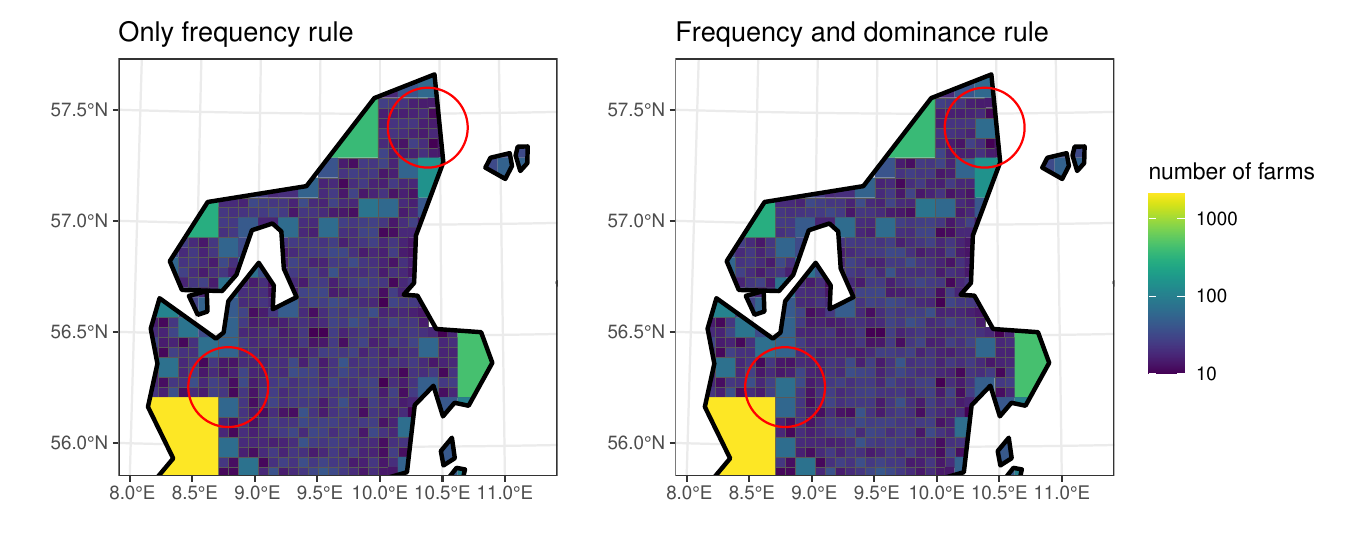}
    \caption{Number of holdings per grid cell for different grid cell sizes for Denmark (based on synthetic data) with different confidentiality rules employed}
    \label{fig:frequencyRulesBoth}
\end{figure}

Figure \ref{fig:frequencyRulesBoth} shows the multi-resolution grids for the synthetic data for Denmark, created with the script above. In the left panel, only the frequency rule was applied, ensuring that all grid cells have at least 10 holdings. The majority of the grid cells (1174) have a resolution of 5 km but there are also 143 with a resolution of 5 km, 29 with 20 km, 8 with 40 km, and 1 with 80 km. It is challenging to see the 5 km grid cells, but we can observe that most of the larger grid cells are on the coastline. Most of the grid cells have 10-50 holdings, but there are 19 grid cells with more than 100 holdings, and one of them with 2078 holdings. 

The panel on the right side of Figure \ref{fig:frequencyRulesBoth} shows the result when the dominance rule is also applied. The difference between the two is small in this case. There are 12 fewer 5 km cells, 1 fewer 10 km cell and 2 more 20 km cells. This difference is caused by some large farm holdings/producers in the grid cells that had to be aggregated. Some differences can be noticed inside the circles, where smaller grid cells have been merged because of the dominance rule.

We can see that several of the grid cells have now become larger, which indicates that the smaller grid cells typically had 1 or 2 very large contributors. There are only 4 grid cells with a 5 km resolution. The majority of grid cells still have a 10 km resolution but there are fewer (213) than in the example above, which uses only a single grid size. There are 55 grid cells with a 20 km resolution, 12 with 40 km, 3 with 80 km and 1 grid cell with a resolution of 160 km. The majority of grid cells still have 10-50 holdings, but now the number of grid cells with more than 100 holdings has increased to 32, where the highest value is still 877.

\subsubsection{UAA and organic UAA}
\begin{figure}[t!]
    \centering
    \includegraphics[width=\textwidth]{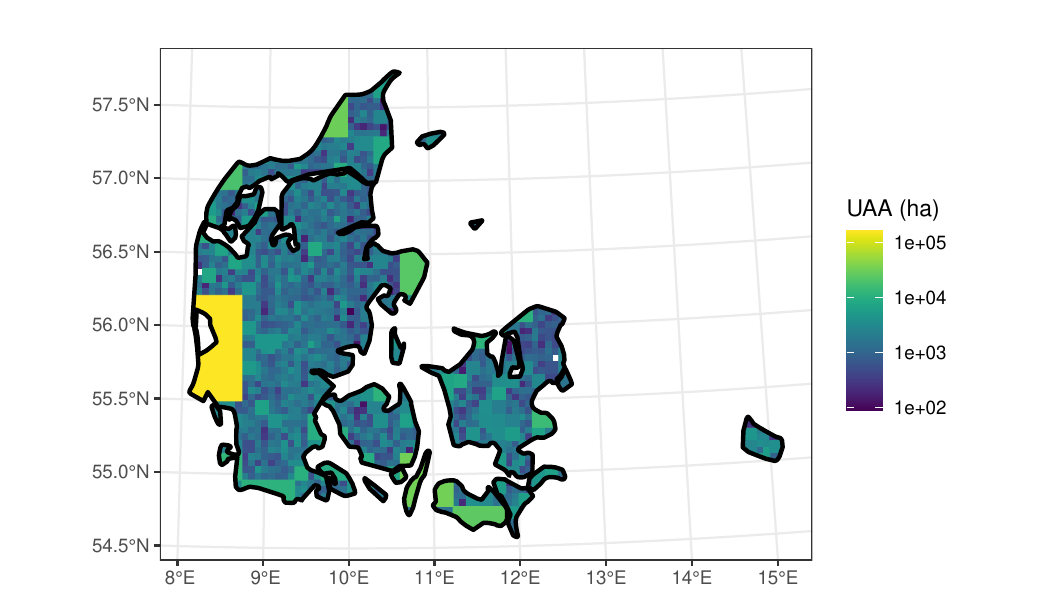}
    \caption{UAA per grid cell for different grid cell sizes for Denmark (synthetic data) }
    \label{fig:uaa}
\end{figure}

\begin{figure}[t!]
    \centering
    \includegraphics[width=\textwidth]{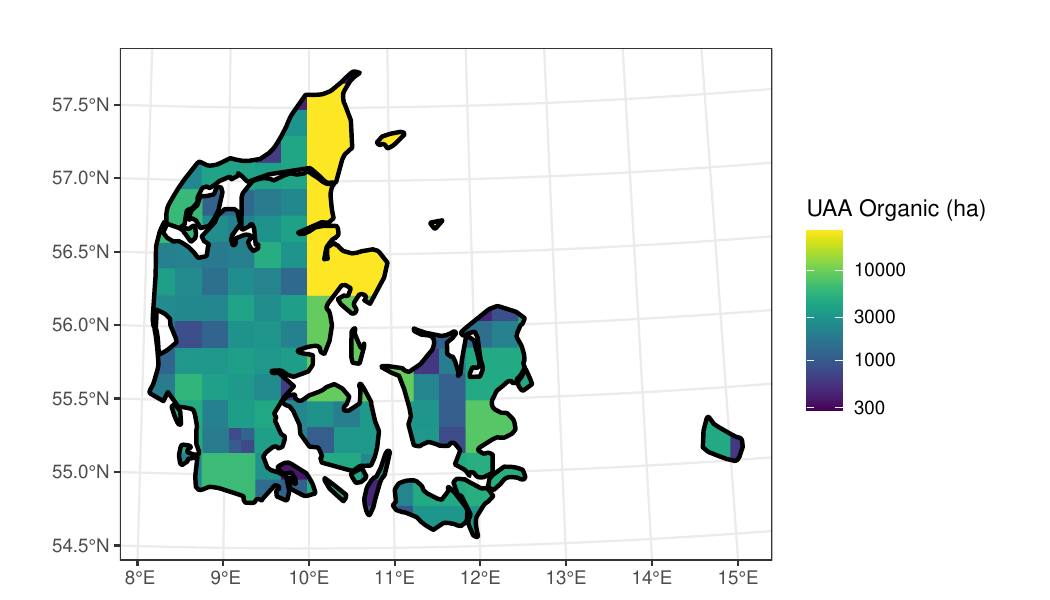}
    \caption{UAA Organic per grid cell for different grid cell sizes for Denmark (synthetic data) }
    \label{fig:uaaorganic}
\end{figure}

The first example provided above focused on the number of holdings, but gridded farm variables will be of more interest to potential users of the data. Two examples are the UAA (already created above) and the organic utilized agricultural area (Organic  UAA, with column name \code{UAAXK0000\_ORG}).  These results are shown in Figures \ref{fig:uaa} and \ref{fig:uaaorganic}, respectively. First, one can notice that the UAA is somewhat dependent on the grid cell size, which is typical for variables that are summed. An alternative would be to present the UAA as UAA/km$^2$ for each grid cell. Second, one can observe that the final grid cells are the same size as in Figure \ref{fig:frequencyRulesBoth} because this is simply another variable produced from the same underlying input data. 

The map of the organic UAA displayed in Figure \ref{fig:uaaorganic} differs, however, the grid cells are generally much larger. This is because there are considerably fewer holdings with organic farming in this synthetic data set. Only one grid cell can be disseminated at 5 km, whereas the majority are 10 km (91) or 20 km (67). Then there are 18, 2 and 1 grid cells of 40 km, 80 km and 160 km, respectively. In total there are 180 grid cells in this map with organic farms.

\subsubsection{Suppressing insignificant grid cells}
The large grid cells on the coast can, in many cases, be an unwanted feature. For many applications, it is usually better to suppress smaller grid cells with a few holdings instead of aggregating them with grid cells that already have a high number of holdings. This can be achieved with a separate argument: \code{suppresslim}.

\begin{figure}[t!]
    \centering
    \includegraphics[width=\textwidth]{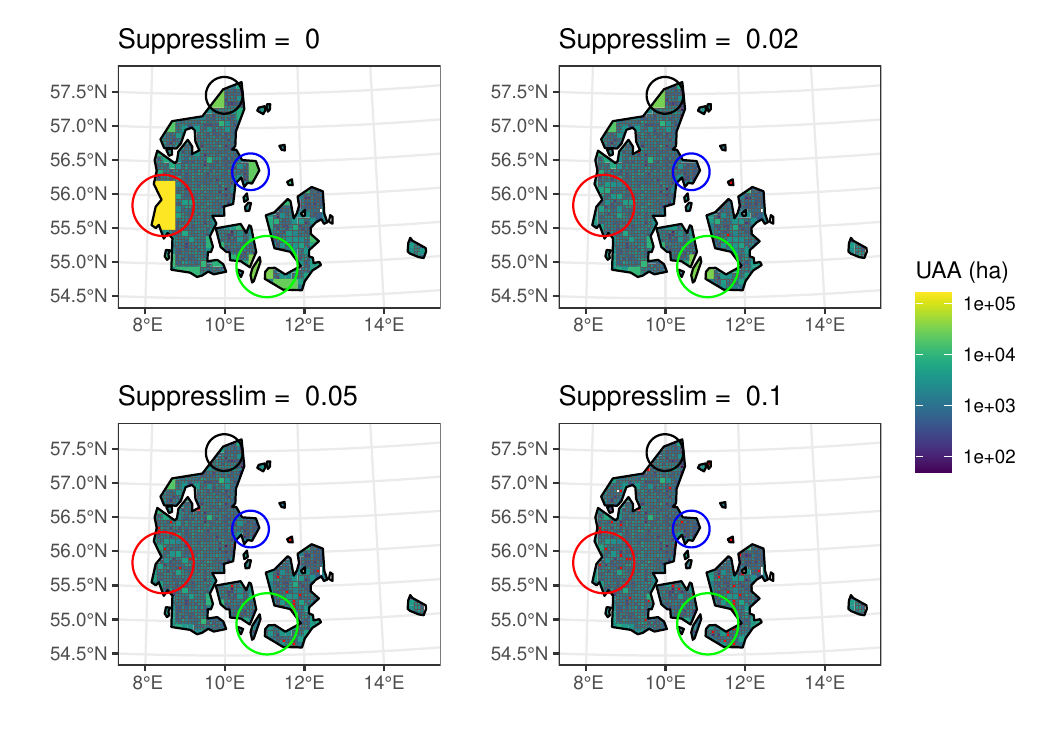}
    \caption{UAA per grid cell for different grid cell sizes for Denmark when different values of $suppresslim$ are provided to the  \fct{multiResGrid} function (synthetic data). Suppressed grid cells are shown in red. The circles highlight regions mentioned in the text.  }
    \label{fig:suppresslim}
\end{figure}

Figure \ref{fig:suppresslim} shows how the grid cell sizes change for UAA, with different values provided to the suppresslim function. This determines the grid cells that should be merged and produces different results as shown in the four panels of Figure \ref{fig:suppresslim}.

When \code{suppresslim = 0} (upper left panel of Figure \ref{fig:suppresslim}), one can see some large grid cells marked with circles that disappear in one of the other panels (as the value provided to the \code{suppresslim} function increases). The ones inside the red and blue circles disappear already with \code{suppresslim = 0.02}. The ones in the black and green circles disappear with \code{suppresslim = 0.05}, and the grid cells within the green circle are further reduced in size for \code{suppresslim = 0.1}. The suppressed grid cells (red squares) are barely visible for the lowest value of \code{suppresslim}, whereas there are considerably more (and larger) grid cells suppressed for the largest value of suppresslim. 

Table \ref{table:suppresslim}  shows the distribution of grid cell sizes for different values of $suppresslim$ provided to the \fct{multiResGrid} function. The numbers in the brackets refer to the number of suppressed grid cells for each grid cell size. There is one value more than in the figure above.

As the value  of $suppresslim$ provided to the \fct{multiResGrid} function increases, the number of large grid cells decreases. For example, the largest grid cell for \code{suppresslim = 0} is 40 km, whereas 20 km is the largest for \code{suppresslim = 0.2}. At the same time, the number of smaller grid cells increases considerably. There are 1774 grid cells of 5 km for \code{suppresslim = 0.2}, whereas there are  1144 for \code{suppresslim = 0}. The number of large grid cells (20 and 40 km) go down from 29 and 8, respectively, to 7 and 0. However, increasing values of suppresslim also leads to suppression of an increasing number of grid cells, in most cases small ones. The total number of suppressed grid cells are 21, 51, 97 and 174, respectively, for the different values in the table. If we look at the percentage of farms and UAA that are not part of the final map, this ranges from 0.3\% - 3.7\% of the total number of farms, and 0.001\% - 2.5\% of the total UAA, with the highest values for \code{suppresslim = 0.2}.

\begin{table}
\centering
\centering
\begin{tabular}[t]{r|r|r|r|r|r}
\hline
\multirow{2}{*}{Resolution (km)} & 
\multicolumn{5}{c}{Suppresslim}\\
 & 0 & 0.02 & 0.05 & 0.1 & 0.2\\
\hline
5 & 1144\hskip 10pt   (0) & 1337 (18) & 1459 (47) & 1591 (90) & 1774 (165)\\
\hline
10 & 151\hskip 10pt   (0) & 178\hskip 10pt   (3) & 170\hskip 10pt   (5) & 147\hskip 10pt   (6) & 101\hskip 10pt   (7)\\
\hline
20 & 29\hskip 10pt   (0) & 18\hskip 10pt   (0) & 15\hskip 10pt   (1) & 10\hskip 10pt   (1) & 7\hskip 10pt   (2)\\
\hline
40 & 8\hskip 10pt   (0) & 5\hskip 10pt   (0) & 2\hskip 10pt   (0) & 1\hskip 10pt   (0) & 0\hskip 10pt   (0)\\
\hline
80 & 1\hskip 10pt   (0) & 0\hskip 10pt   (0) & 0\hskip 10pt   (0) & 0\hskip 10pt   (0) & 0\hskip 10pt   (0)\\
\hline
\end{tabular}
\caption{Distribution of grid cell sizes for different values of suppresslim
            with the number of suppressed grid cells in brackets}
\label{table:suppresslim}
\end{table}

\subsubsection{Demonstrating the need for reliability checks} \label{section:reliability}
In this example, we demonstrate the importance of the reliability checks when gridding survey data as opposed to census data. The reliability check of the estimates does not have an impact on the gridded census data as this covers the entire population. However, this is not the case for the agricultural survey data, which is usually collected in a stratified approach (where the selection depends on a set of criteria that might include geographic location, farm size, economic size, crop types, etc). For each stratum, the surveyed holdings will have a weight assigned depending on the subsampling rate of that stratum. In this case, it is the weighted number of holdings and the weighted values that must pass the confidentiality rules. If a record has a high weight (for example, above 10), the confidentiality rules are immediately met for any grid cell with this record, but the value is not reliable, as it is only based on a single observation. A grid cell estimate can be reliable with considerably less than 10 records, but no fixed limit can be given as this also depends on the population size and the variability within the recorded values.

\begin{figure}[t!]
    \centering
    \includegraphics[width=\textwidth]{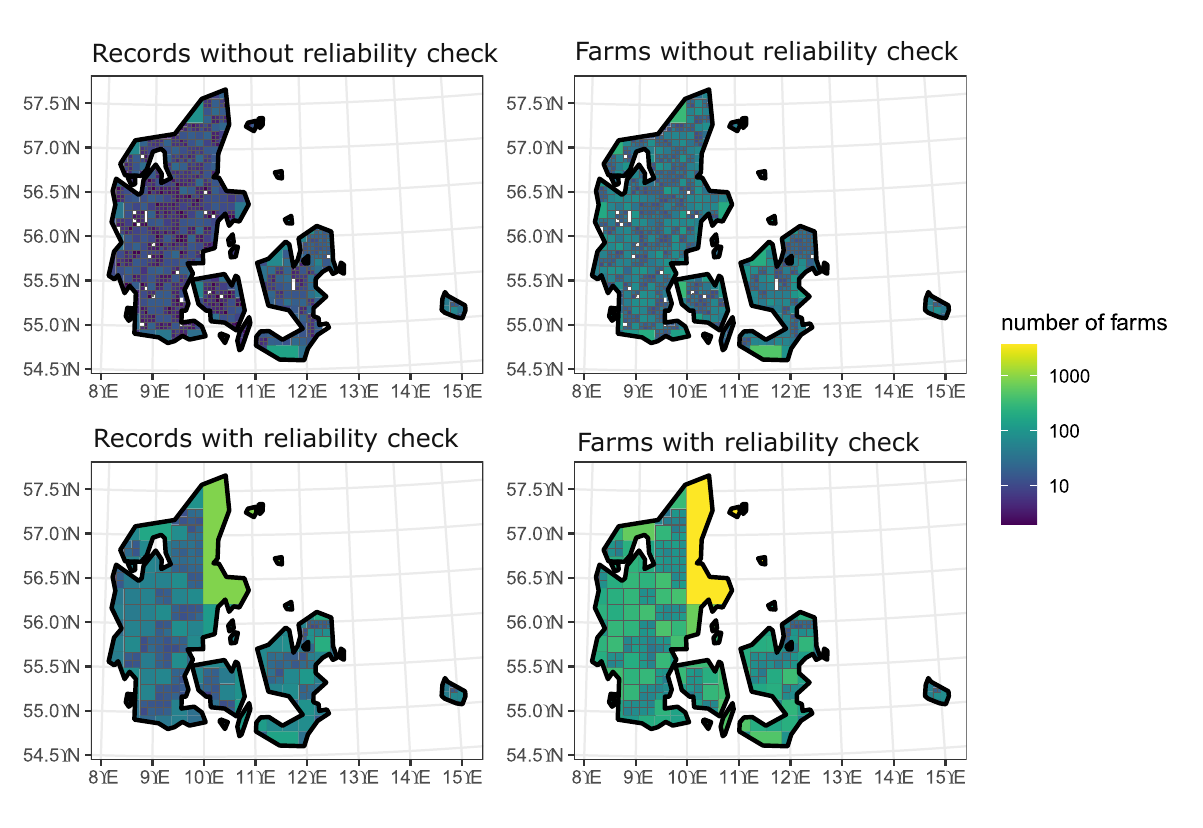}
    \caption{Multi-resolution grid of the number of farms for a synthetic FSS data set for Denmark, without the reliability check on the top and with the reliability check on the bottom. Number of records refers to the actual farms in the survey, whereas the number of farms refers to the weighted number of farms.}
    \label{fig:reliability}
\end{figure}

The results are illustrated in Figure \ref{fig:reliability}, which shows four maps of gridded synthetic agricultural survey data from Denmark, a subset of the data set above. In the top panels of Figure \ref{fig:reliability}, the reliability has not been taken into account, whereas this procedure has been done for the data shown in the bottom panels. The panels to the left show the number of records (the actual number of farms surveyed) whereas the panels to the right show the number of farms according to the survey weights. The tiny grid cells that contain just a few holdings with large enough weights to pass the confidentiality rules have mostly disappeared, producing a smoother and more realistic map. The result is considerably fewer grid cells, based on more records. Only 6 grid cells have less than 10 records, with 3 records as the fewest. Note that the reliability check is integrated into the iterative process that produces the multi-resolution grid, but because it is a computationally intensive process, it is not applied by default.

\subsection{An example of producing a ratio}

\begin{figure}[t!]
    \centering
    \includegraphics[width=\textwidth]{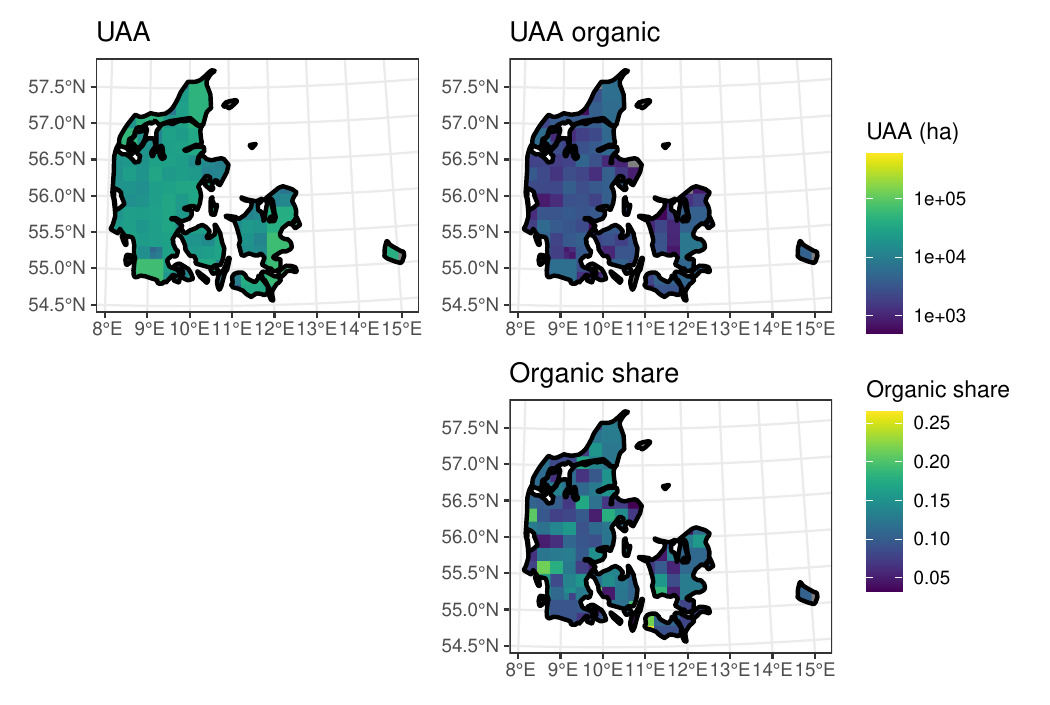}
    \caption{UAA per grid cell for total UAA and organic UAA for different grid cell sizes for Denmark (synthetic data), in addition to the ratio between the two }
    \label{fig:ratio}
\end{figure}

It is only possible to make ratios if the maps have the same resolution. The \code{multiResGrid} function accepts a vector of variable names, and will ensure that the confidentiality rules are respected for all variables for each grid cell. 
Figure \ref{fig:ratio} shows the gridded total UAA and gridded organic UAA in the upper panels, together with the gridded organic share in the lower panel. We can see that the grid cells are the same for both total UAA and organic UAA individually. To produce the ratio or the share of organic UAA, the gridding procedure was done with \code{suppresslim = 0.05}, which resulted in the suppression of two grid cells in the southern part of Denmark and on the island to the east. Using the synthetic data, we can see that the concentration of organic farming is higher in the south of the country, although this pattern may differ when actual data from the agricultural census are used. Note that there is also the function \fct{FSSratio} that can be used if the numerator grid (usually the coarsest) has already been derived and cannot be changed.

\subsection{Producing European wide estimates}
\begin{figure}[t!]
    \centering
    \includegraphics[width=\textwidth]{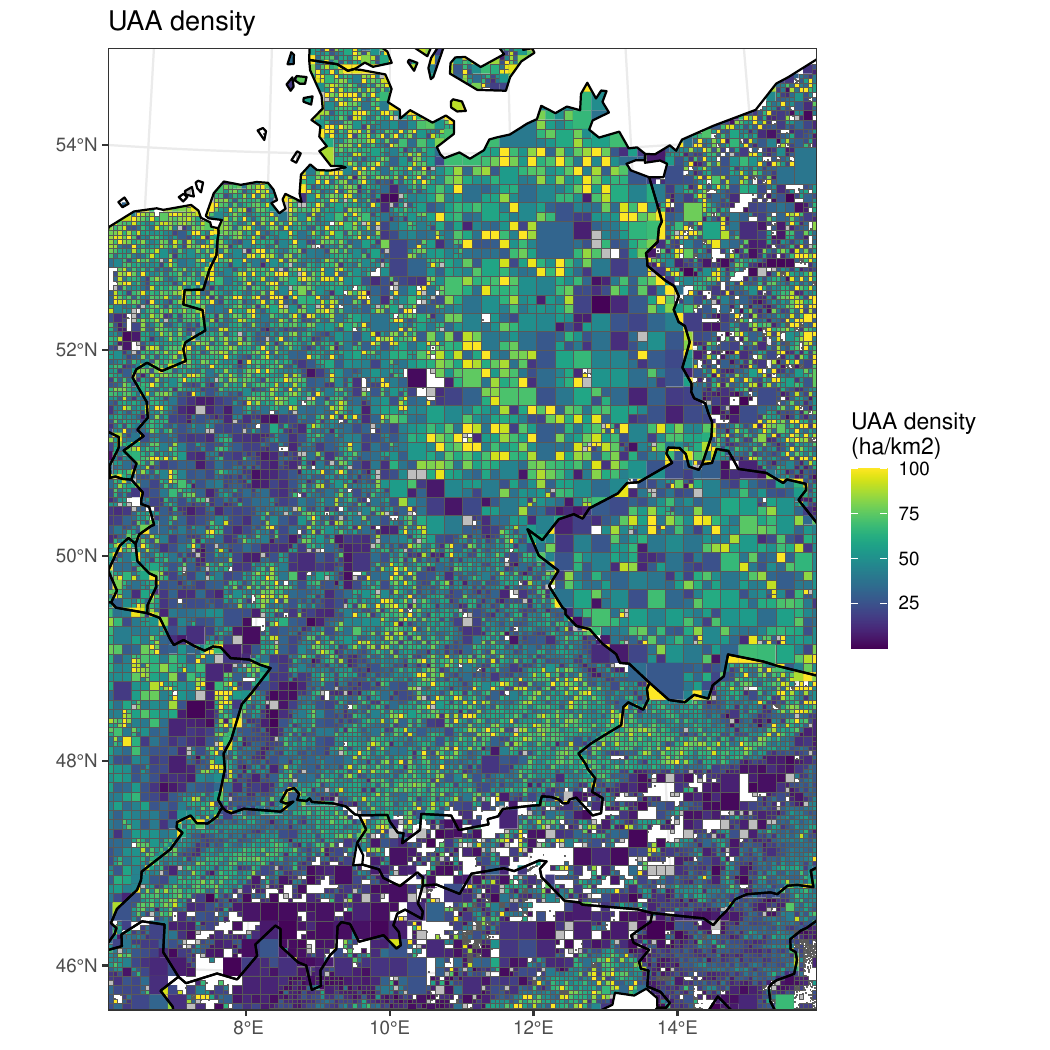}
    \caption{UAA per grid cell for a region in central Europe, based on  2020 FSS data. Suppressed grid cells are shown in gray, white grid cells have no farms. }
    \label{fig:UAA_EU}
\end{figure}

Although we have demonstrated the procedure on national data, the overall aim is to disseminate agricultural variables at a European scale. Since the resolution of a European-wide grid cannot be conveyed with sufficient detail in a figure, here we provide an example of the density of UAA per grid cell in Figure \ref{fig:UAA_EU} for a region in Central Europe, which is based on the 2020 census data from the FSS. It shows the recorded hectares per km\textsuperscript{2}. This can be seen as a surrogate for the percentage of agricultural land in a grid cell (there are 100 ha in one km\textsuperscript{2}). In reality the number of hectares can be higher, especially for smaller grid cells, as some of the agricultural land of a farm might be in different grid cells from the administrative location, used for the gridding. 

We can see how the grid cell sizes vary between different regions and different countries. Mostly areas with high agricultural density also have high resolution grid cells. This is typically the case for large regions in Western Germany, Poland, the north of Austria and the north of Switzerland. Larger grid cells can be found in regions with lower density of agriculture, such as in the Alpine region traversing Austria, Italy and Switzerland, and for example also in large forested areas in Germany (Black Forest) and France (Ballon des Vosges). But there are also large grid cells with high density of agriculture in the former Eastern Germany and in Czechia, most likely due to structural differences that date back to pre-1990 when larger collective farms were present in these countries. 

Table \ref{table:UAA_EU} gives an overview of the number of grid cells with different sizes in this image. We can see that the majority are 1 or 5 km, whereas there are also quite a number of grid cells with a size of 10 and 20 km. There are considerably fewer that are 40 or 80 km.

\begin{table}[H]
\centering
\centering
\begin{tabular}[t]{r|r}
\hline
Resolution (km) &  Number of gridcells\\
\hline
1 & 49705\\
\hline
5 & 74915\\
\hline
10 & 13805\\
\hline
20 & 1853\\
\hline
40 & 279\\
\hline
80 & 97\\
\hline
\end{tabular}
\caption{Distribution of grid cell sizes for the region in Figure \ref{fig:UAA_EU}}
\label{table:UAA_EU}
\end{table}

\subsection{Results from gridding other survey data}
\begin{figure}[t!]
    \centering
    \includegraphics[width=\textwidth]{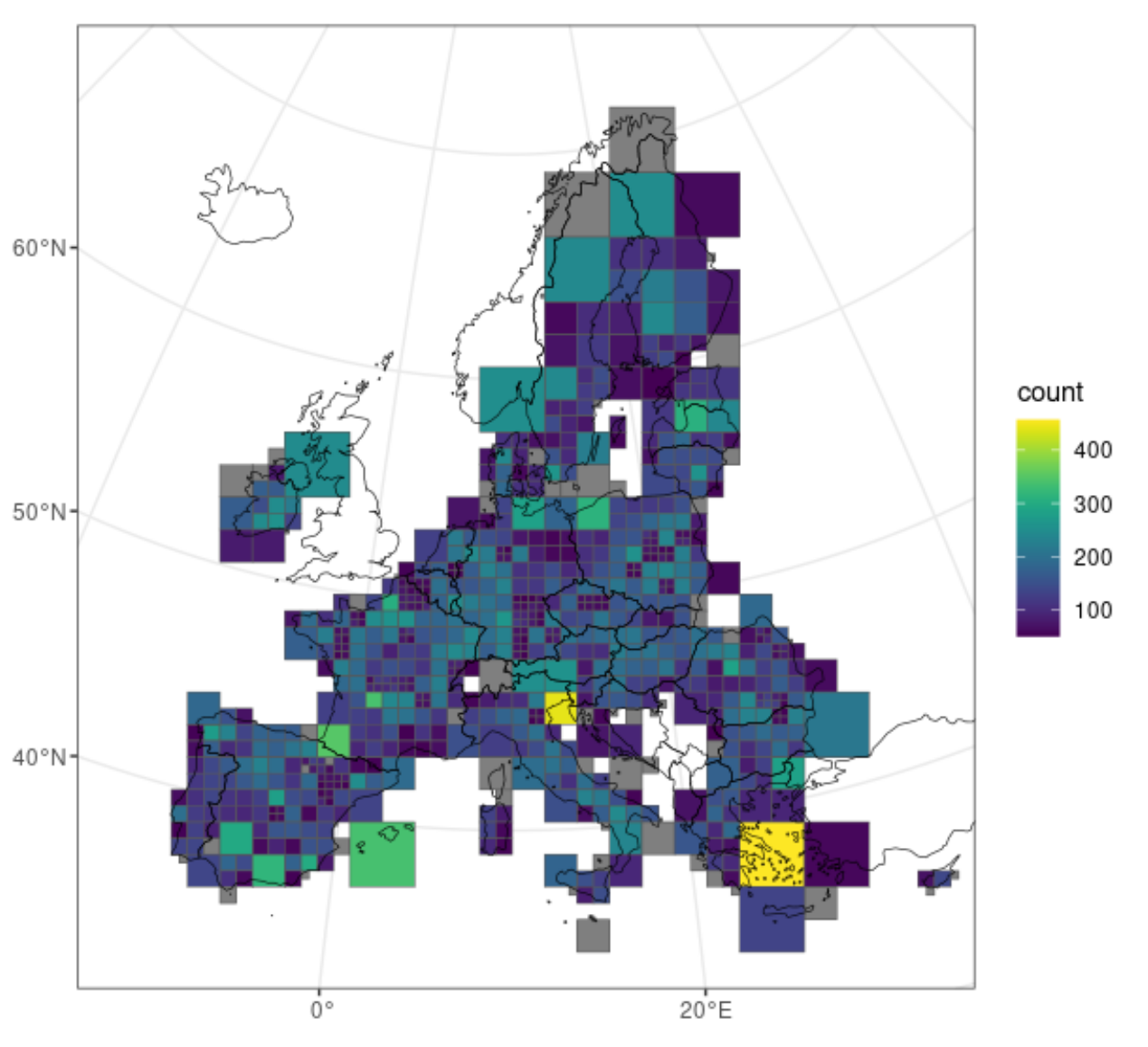}
    \caption{A multi-resolution grid of landscape features from LUCAS data that satisfies a minimum number of observations per grid cell}
    \label{fig:LF}
\end{figure}

Although the method presented here has been developed for agricultural census and survey data, the package can also be applied to survey data for other applications. Figure \ref{fig:LF} shows a grid based on observations of landscape features in the LUCAS survey \citep{Ballin2022}. The landscape feature module from LUCAS consists of 92,633 observations of landscape features in a quadrat. The grid was created so that a minimum of 50 observations is contained within each grid cell to minimize the prediction uncertainty. 

There is some correlation between the observation density and the density of agricultural areas although with some limitations because the sampling was based on a stratified procedure in which homogeneous regions have relatively fewer samples. We can also notice that there are some fairly large grid cells in Figure \ref{fig:LF}. These are typically on the coast, or they partly include countries that were not a part of the survey (which covers only EU countries). Finally, we can observe that there are some suppressed cells in coastal regions, on the borders, and in mountainous regions.

\section{Discussion}

The multi-resolution gridded solution presented here represents a step change in the way that the rich amount of information on the farming sector in Europe, collected by EU Member States and Eurostat in agricultural censuses and surveys, could be released in the future. Using this approach, the information content is maximized and released at the locally highest resolution possible while respecting the confidentiality regulations as specified in EU laws as well as guidelines set by Eurostat and agreed with the Member States. In contrast, other countries outside of the EU are still much stricter in their dissemination of agricultural census data. For example, the United States Department of Agriculture releases data at county level, which is similar to NUTS2 regions in Europe \citep{usda_nass_census_2024}. In Canada, one-third of data were not disclosed in the 2016 agricultural census, which employed suppression of data. For the 2021 Census, Statistics Canada has switched to the use of random tabular adjustment, which makes changes to individual cells to ensure data protection \citep{statistics_canada_guide_2023}. However, the size of the areas for which data are released must be a minimum of 25 square km in area and contain more than 16 farms or the areas are merged with adjacent zones \citep{statistics_canada_census_2023}. Moreover, comparability of the 2021 agricultural census data with previous censuses will be impacted \citep{statistics_canada_guide_2023}. In the UK, the Edinburgh Data and Information Access (EDINA) releases agricultural census data at 2, 5 and 10 km grids \citep{macdonald_counting_2004}. However, with a single grid size, the data are less reliable and/or suppressed in areas where the disclosure requirements are not met \citep{khan_land_2013}. Hence, the suggested approach could be used and adapted by other statistical services that disseminate agricultural census and survey data (such as farm accountancy data) to meet their specific disclosure requirements. Given the versatile and flexible implementation of our approach, the methodology could easily be expanded to other statistical domains where sensitive information on individuals or enterprises is collected, such as population, migration, business and labour force statistics. 

However, there are also limitations with the multi-resolution gridded approach. The examples provided in the paper were for continuous variables. Categorical variables such as farm type, irrigation methods or other gainful activities will require transformation into dummy variables, and these classes will need to be treated individually. Secondly, the reliability check demonstrated in the section \emph{Demonstrating the need for reliability checks} is integrated into the iterative process that produces the multi-resolution grid but it is not applied by default as this process is  computationally time consuming. Finally, creating multi-resolution grids of a ratio requires a different calculation for the estimation of variance in the reliability check, which is currently neglected. This calculation for reliability checking of ratios will be added to the process in the future. 

\section{Conclusions}
In this paper, we presented a method for creating a gridded layer of varying resolutions that maximizes the information content at an aggregated level while respecting confidentiality rules and the recommendations for data disclosure from Eurostat. The method was demonstrated using synthetic FSS data for Denmark on the number of farm holdings, UAA and the number of organic farms. The need for the reliability check was illustrated using a subset of the synthetic data to emulate survey data, which has less data than a full agricultural census and therefore may result in larger grid cells. The method was then demonstrated on the production of ratios, i.e., the organic farming share of UAA, followed by a presentation of a European wide grid of UAA. Finally, the method was applied to a different survey data set, i.e., LUCAS, to demonstrate how a European wide map of landscape features can be generated. 

The method and the R-package includes several features that have not been a part of previous methods for producing multi-resolutiong grids, such as contextual suppression, joint gridding of several variables and the possibility for additional user-defined restrictions. 

The next steps are to apply the method to produce a set of key agricultural indicators from the agricultural census and survey data for Europe, which can be used to better understand agricultural systems across Europe and to identify what drives the adoption of different agricultural practices. The release of grids for analyzing change over time will be more challenging as the multi-resolution grids will need to be spatially consistent if meaningful comparisons are to be made. Methods for ensuring both spatial and temporal consistency will be added in the future. 

This method is only the start of what could be generalized into an on-demand web processing service that would allow users to select the variables of interest and produce multi-resolution grids that do not require large labor resources from Eurostat while respecting all the required confidentiality measures. Such a service could also result in the considerable uptake and use of high resolution European agricultural survey and census data that up to now has only been possible at a coarse and highly aggregated resolution.

\section*{Acknowledgments}
The authors appreciate very much the support, collaboration and agreement from the colleagues of Statistics Denmark to share the synthetic data file of agricultural census 2020 with the public. Additionally, we thank colleagues from Eurostat unit of Methodology and Innovation in official statistics for their useful and constructive comments received during the early phase of the research project. 
\newpage

\bibliographystyle{elsarticle-harv} 
\bibliography{refs}

\newpage

\begin{appendix}

\section{Synthetic public file for Denmark (DK)}
\label{app:synth}
\begin{table}[!htbp] \centering 
  \caption{Summary statistics by type of data file} 
  \label{tab:syn} 
\begin{tabular}{@{\extracolsep{5pt}} cccc} 
\\[-1.8ex]\hline 
\hline \\[-1.8ex] 
 & Real & Synthetic & p \\ 
\hline \\[-1.8ex] 
Sample size & 37088 & 37088 &  \\ 
UAA & 70.91 (141.86) & 70.48 (143.76) &  0.678 \\ 
UAAXK0000\_ORG &  8.26 (53.86) &  7.84 (50.42) &  0.273 \\ 
\hline \\[-1.8ex] 
\end{tabular} 
\end{table} 
\newpage
\begin{figure}[t!]
    \centering
\includegraphics[width=\textwidth]{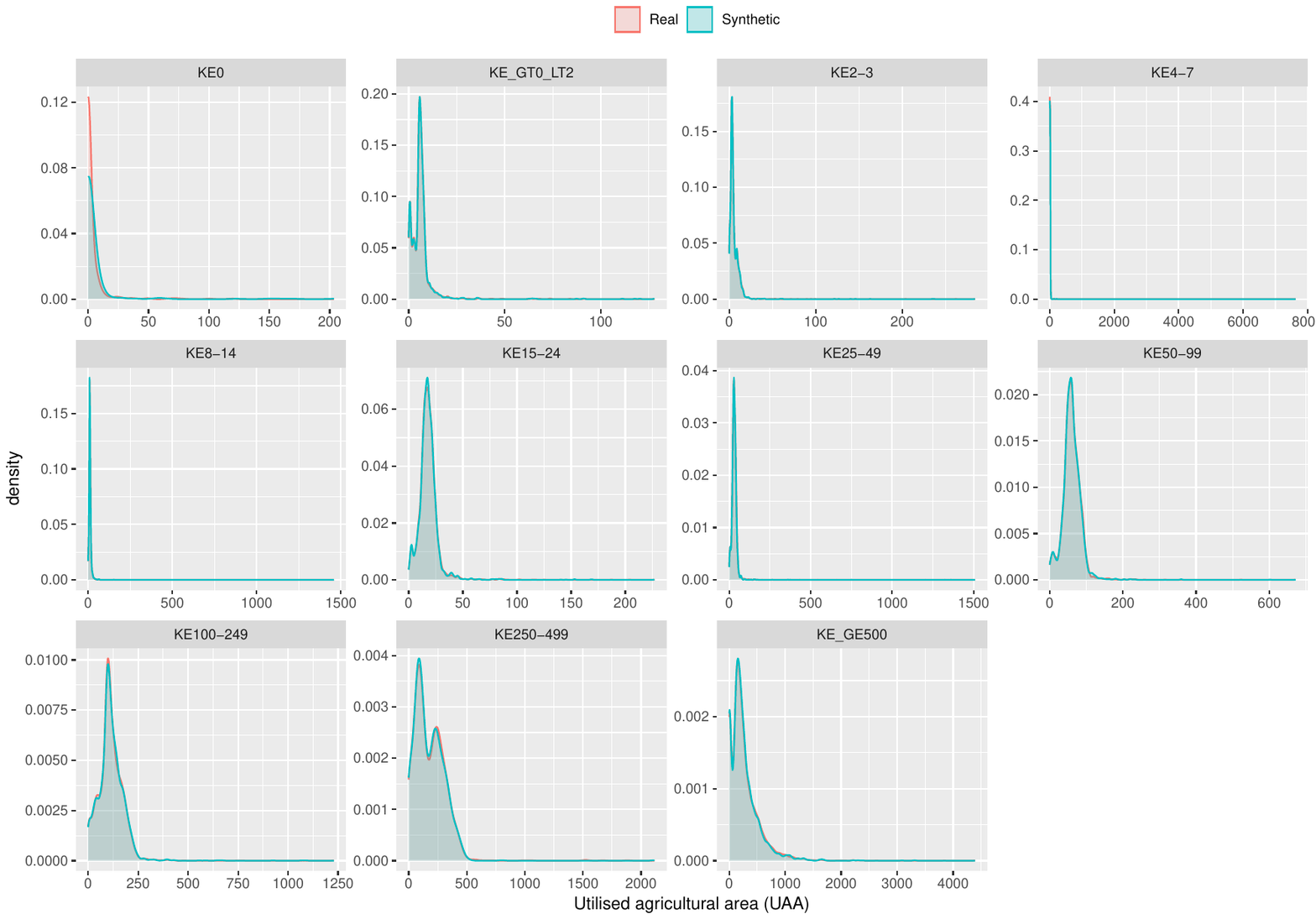}
    \caption{Distribution of utilised agricultural area by economic size classes}
    \label{fig:syn_uaa}
\end{figure}

\begin{figure}[t!]
    \centering
\includegraphics[width=\textwidth]{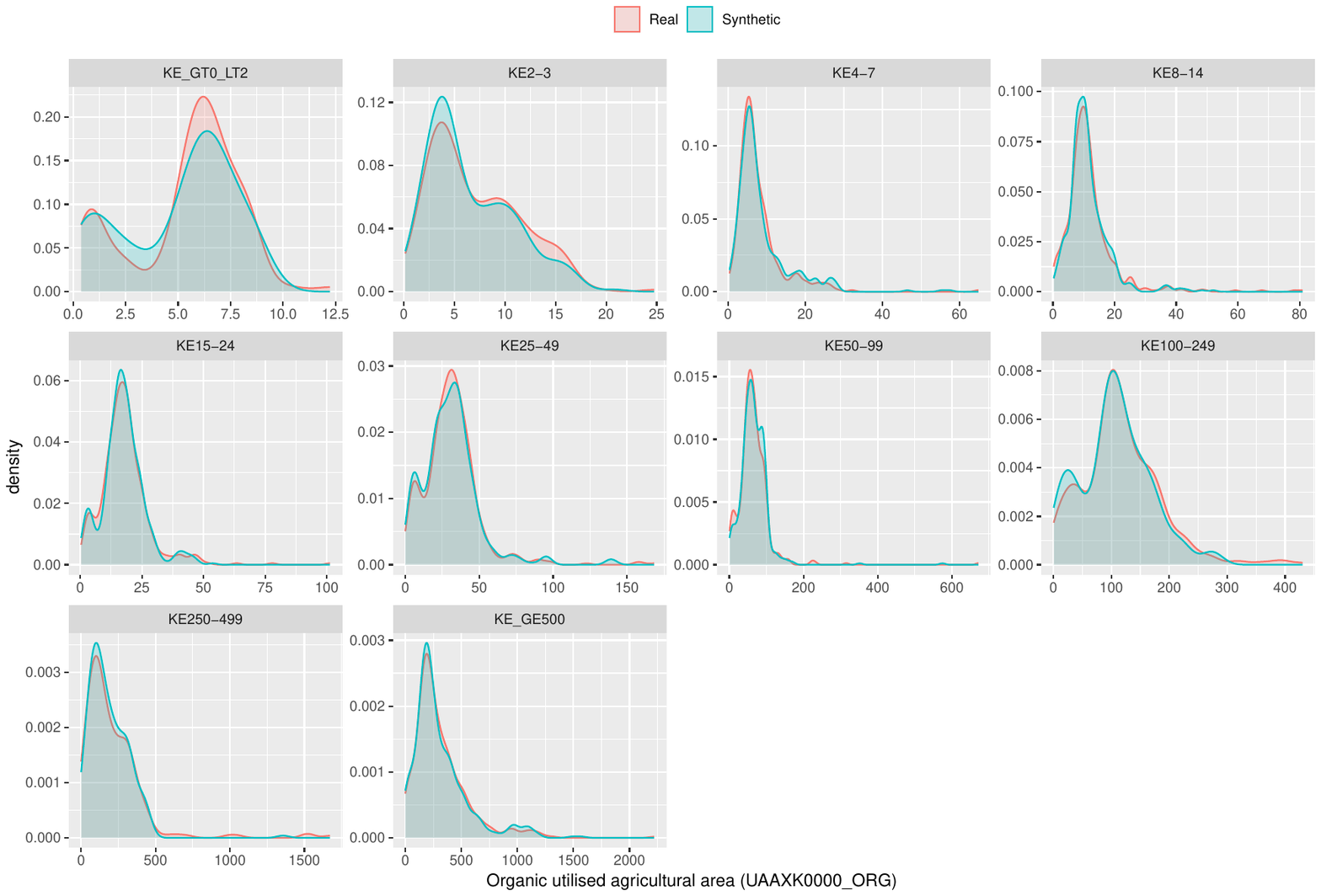}
    \caption{Distribution of organic utilised agricultural area by economic size classes}
    \label{fig:syn_uaax}
\end{figure}

\end{appendix}


\end{document}